\documentclass[reprint,superscriptaddress,preprintnumbers,longbibliography,
amsmath,amssymb,aps,floatfix,pra,twocolumn, tightenlines %linenumbers
]{revtex4-2}
\usepackage{subfigure}
\usepackage{mathrsfs}
\usepackage[T1]{fontenc}
\usepackage{graphicx}% Include figure files
\usepackage{dcolumn}% Align table columns on decimal point
\usepackage{bm}% bold math
\usepackage{color}
\usepackage{times}
\usepackage[colorlinks=true,
citecolor=blue,
linkcolor=magenta,
anchorcolor=black,
urlcolor=magenta]{hyperref}
\usepackage{siunitx}
\usepackage{overpic}
\usepackage{braket}
\usepackage{changes}
\usepackage{amssymb}
\newcommand{\mm}[1]{\mathrm{#1}}

\usepackage{float}

\begin{document}

	\title{Prethermalization by Random Multipolar Driving on a 78-Qubit Superconducting Processor}
	
	\author{Zheng-He Liu}
	\thanks{These authors contributed equally to this work.}
	\affiliation{Beijing National Laboratory for Condensed Matter Physics, Institute of Physics, Chinese Academy of Sciences, Beijing 100190, China}
	\affiliation{School of Physical Sciences, University of Chinese Academy of Sciences, Beijing 100049, China}
	
	\author{Yu Liu}
	\thanks{These authors contributed equally to this work.}
	\affiliation{Beijing National Laboratory for Condensed Matter Physics, Institute of Physics, Chinese Academy of Sciences, Beijing 100190, China}
	\affiliation{School of Physical Sciences, University of Chinese Academy of Sciences, Beijing 100049, China}
	
	\author{Gui-Han Liang}
	\thanks{These authors contributed equally to this work.}
	\affiliation{Beijing National Laboratory for Condensed Matter Physics, Institute of Physics, Chinese Academy of Sciences, Beijing 100190, China}
	\affiliation{School of Physical Sciences, University of Chinese Academy of Sciences, Beijing 100049, China}
	
	\author{Cheng-Lin Deng}
	\thanks{These authors contributed equally to this work.}
	\affiliation{Beijing National Laboratory for Condensed Matter Physics, Institute of Physics, Chinese Academy of Sciences, Beijing 100190, China}
	\affiliation{School of Physical Sciences, University of Chinese Academy of Sciences, Beijing 100049, China}
	
	\author{Keyang Chen}
	\affiliation{School of Physical Sciences, University of Chinese Academy of Sciences, Beijing 100049, China}
	\affiliation{CAS Key Laboratory for Theoretical Physics, Institute of Theoretical Physics, Chinese Academy of Sciences, Beijing 100190, China}
	
	\author{Yun-Hao Shi}
	\affiliation{Beijing National Laboratory for Condensed Matter Physics, Institute of Physics, Chinese Academy of Sciences, Beijing 100190, China}
	
	\author{Tian-Ming Li}
	\affiliation{Beijing National Laboratory for Condensed Matter Physics, Institute of Physics, Chinese Academy of Sciences, Beijing 100190, China}
	\affiliation{School of Physical Sciences, University of Chinese Academy of Sciences, Beijing 100049, China}
	
	\author{Lv Zhang}
	\affiliation{Beijing National Laboratory for Condensed Matter Physics, Institute of Physics, Chinese Academy of Sciences, Beijing 100190, China}
	\affiliation{School of Physical Sciences, University of Chinese Academy of Sciences, Beijing 100049, China}
	
	\author{Bing-Jie Chen}	
	\affiliation{Beijing National Laboratory for Condensed Matter Physics, Institute of Physics, Chinese Academy of Sciences, Beijing 100190, China}
	\affiliation{School of Physical Sciences, University of Chinese Academy of Sciences, Beijing 100049, China}
	
	\author{Cai-Ping Fang}
	\affiliation{Beijing National Laboratory for Condensed Matter Physics, Institute of Physics, Chinese Academy of Sciences, Beijing 100190, China}
	\affiliation{School of Physical Sciences, University of Chinese Academy of Sciences, Beijing 100049, China}

	\author{Da'er Feng}
	\affiliation{Beijing National Laboratory for Condensed Matter Physics, Institute of Physics, Chinese Academy of Sciences, Beijing 100190, China}
	\affiliation{School of Physical Sciences, University of Chinese Academy of Sciences, Beijing 100049, China}
	
	\author{Xu-Yang Gu}
	\affiliation{Beijing National Laboratory for Condensed Matter Physics, Institute of Physics, Chinese Academy of Sciences, Beijing 100190, China}
	\affiliation{School of Physical Sciences, University of Chinese Academy of Sciences, Beijing 100049, China}
	
	\author{Yang He}
	\affiliation{Beijing National Laboratory for Condensed Matter Physics, Institute of Physics, Chinese Academy of Sciences, Beijing 100190, China}
	\affiliation{School of Physical Sciences, University of Chinese Academy of Sciences, Beijing 100049, China}
	
	\author{Kaixuan Huang}
	\affiliation{Beijing Key Laboratory of Fault-Tolerant Quantum Computing, Beijing Academy of Quantum Information Sciences, Beijing 100193, China}
	
	\author{Hao Li}
	\affiliation{Beijing Key Laboratory of Fault-Tolerant Quantum Computing, Beijing Academy of Quantum Information Sciences, Beijing 100193, China}
	
	\author{Hao-Tian Liu}
	\affiliation{Beijing National Laboratory for Condensed Matter Physics, Institute of Physics, Chinese Academy of Sciences, Beijing 100190, China}
	\affiliation{School of Physical Sciences, University of Chinese Academy of Sciences, Beijing 100049, China}
	
	\author{Li Li}
	\affiliation{Beijing National Laboratory for Condensed Matter Physics, Institute of Physics, Chinese Academy of Sciences, Beijing 100190, China}
	\affiliation{School of Physical Sciences, University of Chinese Academy of Sciences, Beijing 100049, China}
	
	\author{Zheng-Yang Mei}
	\affiliation{Beijing National Laboratory for Condensed Matter Physics, Institute of Physics, Chinese Academy of Sciences, Beijing 100190, China}
	\affiliation{School of Physical Sciences, University of Chinese Academy of Sciences, Beijing 100049, China}
	
	\author{Zhen-Yu Peng}
	\affiliation{Beijing National Laboratory for Condensed Matter Physics, Institute of Physics, Chinese Academy of Sciences, Beijing 100190, China}
	\affiliation{School of Physical Sciences, University of Chinese Academy of Sciences, Beijing 100049, China}
	
	\author{Jia-Cheng Song}
	\affiliation{Beijing National Laboratory for Condensed Matter Physics, Institute of Physics, Chinese Academy of Sciences, Beijing 100190, China}
	\affiliation{School of Physical Sciences, University of Chinese Academy of Sciences, Beijing 100049, China}
	
	\author{Ming-Chuan Wang}
	\affiliation{Beijing National Laboratory for Condensed Matter Physics, Institute of Physics, Chinese Academy of Sciences, Beijing 100190, China}
	\affiliation{School of Physical Sciences, University of Chinese Academy of Sciences, Beijing 100049, China}
	
	\author{Shuai-Li Wang}
	\affiliation{Beijing National Laboratory for Condensed Matter Physics, Institute of Physics, Chinese Academy of Sciences, Beijing 100190, China}
	\affiliation{School of Physical Sciences, University of Chinese Academy of Sciences, Beijing 100049, China}
	
	\author{Ziting Wang}
	\affiliation{Beijing Key Laboratory of Fault-Tolerant Quantum Computing, Beijing Academy of Quantum Information Sciences, Beijing 100193, China}
	
	\author{Yongxi Xiao}
	\affiliation{Beijing National Laboratory for Condensed Matter Physics, Institute of Physics, Chinese Academy of Sciences, Beijing 100190, China}
	\affiliation{School of Physical Sciences, University of Chinese Academy of Sciences, Beijing 100049, China}
	
	\author{Minke Xu}
	\affiliation{Beijing National Laboratory for Condensed Matter Physics, Institute of Physics, Chinese Academy of Sciences, Beijing 100190, China}
	\affiliation{School of Physical Sciences, University of Chinese Academy of Sciences, Beijing 100049, China}
	
	\author{Yue-Shan Xu}
	\affiliation{Beijing Key Laboratory of Fault-Tolerant Quantum Computing, Beijing Academy of Quantum Information Sciences, Beijing 100193, China}
	
	\author{Yu Yan}
	\affiliation{Beijing National Laboratory for Condensed Matter Physics, Institute of Physics, Chinese Academy of Sciences, Beijing 100190, China}
	\affiliation{School of Physics, Northwest University, Xi’an 710127, China}
	
	\author{Yi-Han Yu}
	\affiliation{Beijing National Laboratory for Condensed Matter Physics, Institute of Physics, Chinese Academy of Sciences, Beijing 100190, China}
	\affiliation{School of Physical Sciences, University of Chinese Academy of Sciences, Beijing 100049, China}
	
	\author{Wei-Ping Yuan}
	\affiliation{Beijing National Laboratory for Condensed Matter Physics, Institute of Physics, Chinese Academy of Sciences, Beijing 100190, China}
	\affiliation{School of Physical Sciences, University of Chinese Academy of Sciences, Beijing 100049, China}
	
	\author{Jia-Chi Zhang}
	\affiliation{Beijing National Laboratory for Condensed Matter Physics, Institute of Physics, Chinese Academy of Sciences, Beijing 100190, China}
	\affiliation{School of Physical Sciences, University of Chinese Academy of Sciences, Beijing 100049, China}
	
	\author{Jun-Jie Zhao}
	\affiliation{Beijing National Laboratory for Condensed Matter Physics, Institute of Physics, Chinese Academy of Sciences, Beijing 100190, China}
	\affiliation{School of Physical Sciences, University of Chinese Academy of Sciences, Beijing 100049, China}
	
	\author{Kui Zhao}
	\affiliation{Beijing Key Laboratory of Fault-Tolerant Quantum Computing, Beijing Academy of Quantum Information Sciences, Beijing 100193, China}
	
	\author{Si-Yun Zhou}
	\affiliation{Beijing National Laboratory for Condensed Matter Physics, Institute of Physics, Chinese Academy of Sciences, Beijing 100190, China}
	\affiliation{School of Physical Sciences, University of Chinese Academy of Sciences, Beijing 100049, China}

    \author{Zheng-An Wang}
    \affiliation{Beijing Key Laboratory of Fault-Tolerant Quantum Computing, Beijing Academy of Quantum Information Sciences, Beijing 100193, China}

	\author{Xiaohui Song}
	\affiliation{Beijing National Laboratory for Condensed Matter Physics, Institute of Physics, Chinese Academy of Sciences, Beijing 100190, China}
	
	\author{Ye Tian}
	\affiliation{Beijing National Laboratory for Condensed Matter Physics, Institute of Physics, Chinese Academy of Sciences, Beijing 100190, China}
	
	\author{Florian Mintert}
	\affiliation{\small Blackett Laboratory, Imperial College London, London SW7 2AZ, United Kingdom}
	\affiliation{Helmholtz-Zentrum Dresden-Rossendorf, Bautzner Landstra{\ss}e 400, 01328 Dresden, Germany}
	
	\author{Johannes Knolle  }
	\affiliation{Department of Physics TQM, Technische Universit{\"a}t M{\"u}nchen, James-Franck-Stra{\ss}e 1, D-85748 Garching, Germany}
	\affiliation{Munich Center for Quantum Science and Technology (MCQST), 80799 Munich, Germany}
	\affiliation{\small Blackett Laboratory, Imperial College London, London SW7 2AZ, United Kingdom}
	
	\author{Roderich Moessner}
	\affiliation{\small Max-Planck-Institut f{\"u}r Physik komplexer Systeme, N{\"o}thnitzer Stra{\ss}e 38, 01187 Dresden, Germany}
	
	\author{Yu-Ran Zhang}
	\affiliation{School of Physics and Optoelectronics, South China University of Technology, Guangzhou 510640, China}
	
	\author{Pan Zhang}
	\affiliation{School of Physical Sciences, University of Chinese Academy of Sciences, Beijing 100049, China}
	\affiliation{CAS Key Laboratory for Theoretical Physics, Institute of Theoretical Physics, Chinese Academy of Sciences, Beijing 100190, China}
	
	\author{Zhongcheng Xiang}
	\email{zcxiang@iphy.ac.cn}
	\affiliation{Beijing National Laboratory for Condensed Matter Physics, Institute of Physics, Chinese Academy of Sciences, Beijing 100190, China}
	\affiliation{School of Physical Sciences, University of Chinese Academy of Sciences, Beijing 100049, China}
	\affiliation{Hefei National Laboratory, Hefei 230088, China}
	
	\author{Dongning Zheng}
	\affiliation{Beijing National Laboratory for Condensed Matter Physics, Institute of Physics, Chinese Academy of Sciences, Beijing 100190, China}
	\affiliation{School of Physical Sciences, University of Chinese Academy of Sciences, Beijing 100049, China}
	\affiliation{Hefei National Laboratory, Hefei 230088, China}
	\affiliation{Songshan Lake Materials Laboratory, Dongguan, Guangdong 523808, China}

	\author{Kai Xu}
	\email{kaixu@iphy.ac.cn}
	\affiliation{Beijing National Laboratory for Condensed Matter Physics, Institute of Physics, Chinese Academy of Sciences, Beijing 100190, China}
	\affiliation{School of Physical Sciences, University of Chinese Academy of Sciences, Beijing 100049, China}
    \affiliation{Beijing Key Laboratory of Fault-Tolerant Quantum Computing, Beijing Academy of Quantum Information Sciences, Beijing 100193, China}
	\affiliation{Hefei National Laboratory, Hefei 230088, China}
	\affiliation{Songshan Lake Materials Laboratory, Dongguan, Guangdong 523808, China}

	\author{Hongzheng Zhao}
	\email{hzhao@pku.edu.cn}
	\affiliation{State Key Laboratory of Artificial Microstructure and Mesoscopic Physics, School of Physics, Peking University, Beijing 100871, China}

	\author{Heng Fan}
	\email{hfan@iphy.ac.cn}
	\affiliation{Beijing National Laboratory for Condensed Matter Physics, Institute of Physics, Chinese Academy of Sciences, Beijing 100190, China}
	\affiliation{School of Physical Sciences, University of Chinese Academy of Sciences, Beijing 100049, China}
    \affiliation{Beijing Key Laboratory of Fault-Tolerant Quantum Computing, Beijing Academy of Quantum Information Sciences, Beijing 100193, China}
	\affiliation{Hefei National Laboratory, Hefei 230088, China}
	\affiliation{Songshan Lake Materials Laboratory, Dongguan, Guangdong 523808, China}

% 	\date{\today}
	
	\begin{abstract}
		Time-dependent drives hold the promise of realizing non-equilibrium many-body phenomena that are absent in undriven systems. Yet, drive-induced heating normally destabilizes the systems, which can be parametrically suppressed in the high-frequency regime by using periodic (Floquet) drives. It remains largely unknown to what extent highly controllable quantum simulators can suppress heating in non-periodically driven systems. Using the 78-qubit superconducting quantum processor, \textit{Chuang-tzu 2.0}, we report the experimental observation of long-lived prethermal phases in many-body systems with tunable heating rates, driven by structured random protocols, characterized by $n$-multipolar temporal correlations. By measuring both the particle imbalance and subsystem entanglement entropy, we monitor the entire heating process over 1,000 driving cycles and observe the existence of the prethermal plateau. The prethermal lifetime is `doubly tunable': one way by driving frequency, the other by multipolar order; it
		grows algebraically with the frequency with the universal scaling exponent $2n{+}1$. Using quantum state tomography on different subsystems, we demonstrate a non-uniform spatial entanglement distribution and observe a crossover from area-law to volume-law entanglement scaling.
		With 78 qubits and 137 couplers in a 2D configuration, the entire far-from-equilibrium heating dynamics are beyond the reach of simulation using tensor-network numerical techniques. Our work highlights superconducting quantum processors as a powerful platform for exploring universal scaling laws and non-equilibrium phases of matter in driven systems in regimes where classical simulation faces formidable challenges.
	\end{abstract}
	
	\maketitle
	
% 	Driven quantum systems can arise exotic phenomena that are absent in their static counterparts~\cite{Eisert2015,Fu2024}. In particular, 
	Periodically driven (Floquet) systems can host novel far-from-equilibrium phenomena that are absent in thermal equilibrium~\cite{Eisert2015,Eckardt2017,Ueda2020}. Prominent examples include the discrete-time crystals~\cite{PhaseStructure2016,Else2016, Choi2017, Yao2017, Rovny2018, Randall2021, Zhang2022, Zaletel2023}, Floquet topological matter~\cite{Aidelsburger2013,Rudner2013, Titum2016, Rudner2020} and dynamical phase transitions~\cite{Bastidas2012,Yang2019,Naji2022}. Periodic drives have also been widely 
    employed for Floquet engineering of many-body interactions~\cite{schweizer2019floquet,Geier2021,MagnonBoundStates2023} and mitigating environment-induced decoherence~\cite{Viebahn2021, Bai2023},
	serving as a robust and versatile approach to stabilize and control modern quantum simulators~\cite{Weitenberg2021, maskara2021discrete,Zhao2022Probing, Shi2023,onedimensionalanyons2024,wang_realization_2024,Rosen2024,Fu2024,Liu_interplay2025}.
     Explorations of non-periodic driving have surged in recent years, with rich discoveries of non-equilibrium phenomena beyond the Floquet lore~\cite{Nandy2017,Dumitrescu2018,Boyers2020,Crowley2020,ZhaoRandom2021,TimmsQuantizedFloquet2021,long2021nonadiabatic,Zhao2022Suppression,Timeperiodicity2022,Moon2024,He2024,schmid2024self,Hilbert-Space-Ergodicity2024,Prethermalization2024,pilatowsky2025critically,wu2025geometricquantumdriveshyperbolically}. For instance, quasi-periodic and structured random drives can lead to the appearance of discrete-time quasi-crystals~\cite{Dumitrescu2018,zhao2019floquet,else2020long,Time-Quasicrystals2025} and time rondeau crystals~\cite{Moon2024,ma2025stabletimerondeaucrystals}, notably enriching the possible forms of temporal order in non-equilibrium settings.

    \begin{figure*}[t]
        \centering
    	\includegraphics[width=0.95\linewidth]{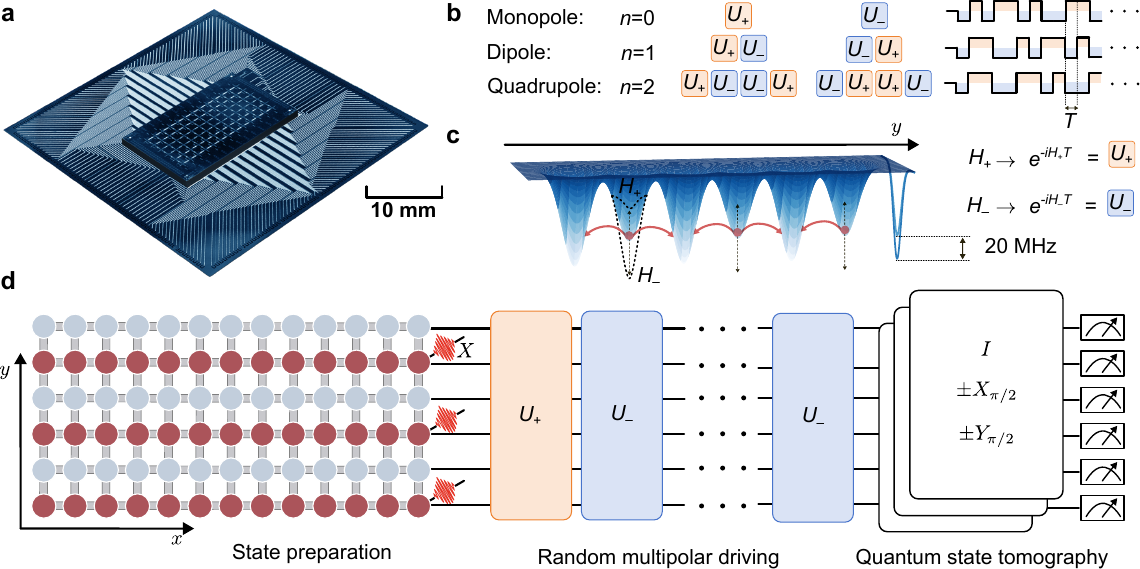}
    	\caption{\textbf{Quantum processor and experimental scheme.} \textbf{a}, Optical micrograph of the 78-qubit superconducting processor, \textit{Chuang-tzu 2.0}. The processor is designed as a 6 $\times$ 13 square lattice, comprising 78 qubits interconnected by 137 couplers that link all adjacent qubits. \textbf{b}, Schematic diagram of the random multipolar driving (RMD) protocol, characterized by the temporal multipolar order $n$. The 0-RMD  is constructed by randomly selecting elements from the two elementary operators $\{\hat{U}_+,\hat{U}_-\}$, and $n$-RMD sequence consists of a random selection of two $n$-multipoles, which are obtained by anti-aligning two $(n-1)$-th order operators.
    	\textbf{c}, Elementary operators are generated as $\hat{U}_{\pm}=\exp\{-\mm{i}\hat{H}_{\pm}T\}$, with a driving period $T$, where $\hat{H}_{\pm}$ differs in the staggered potential in $y$-direction. \textbf{d}, Experimental procedure: First, we initialize the system in a density wave state, where even sites along the $y$-axis are occupied. Next, we implement the RMD protocol that destabilizes the density wave order, and the system heats up. Finally, we use multiqubit quantum state tomography to determine the entanglement entropy and perform particle number measurement to characterize the non-equilibrium time evolution.}
	\label{fig1}
    \end{figure*}
    
    Due to the lack of energy conservation, generic time-dependent many-body systems are inherently susceptible to heating, eventually ending in a featureless infinite-temperature state~\cite{lazarides2014equilibrium,DAlessio2014}, where the subsystem entanglement entropy also reaches its maximum -- the Page value~\cite{Page1993}. This heating effect thus poses a fundamental challenge for utilizing large-scale quantum simulators and stabilizing sought-after phases, especially over long timescales. In Floquet systems, heating can be suppressed by many-body localization (MBL), induced via efficiently strong spatial disorder~\cite{Bordia2017,Abanin2019,Decker2020}. 
    In clean systems, on the other hand, heating can also be exponentially suppressed by using high-frequency drives~\cite{ExponentiallySlowHeating2015,RigorousBound2016}, leading to the transient but long-lived prethermal regime before the eventual heat death~\cite{Berges2004,Gring2012,Mallayya2019,RubioAbadal2020, Peng2021,Beatrez2023,He2023, Ho2023,Harkins2024}. 
    In contrast, stabilizing non-periodically driven systems is a notoriously difficult task, especially when the driving protocol involves temporal randomness. This typically opens deleterious energy absorption channels, which even MBL cannot prevent, and thus heating occurs swiftly.
    
    In this work, we experimentally demonstrate the existence of a long-lived, doubly tunable prethermal regime on a quantum simulator driven by random but structured protocols, with a universal degree of tunability in the heating rate. We use a superconducting quantum processor, \textit{Chuang-tzu 2.0} as shown in Fig.~\ref{fig1}, that involves 78 qubits and 137 couplers.
    Leveraging the precise control and flexibility of this device, we accurately implement stable long-term drives and perform large-scale analog quantum simulations of 2D hard-core Bose-Hubbard system. 
    
    Furthermore, we implement a family of structured random protocols, known as random multipolar driving (RMD)~\cite{ZhaoRandom2021,Mori2021}. As illustrated in Fig.~\ref{fig1}\textbf{b}, the protocol involves two elementary evolution operators $\hat{U}_+$ and $\hat{U}_-$, generated by two Hamiltonians $\hat{H}_{\pm}$ that differ in the chemical potential along the y-axis, see Fig.~\ref{fig1}\textbf{b} and \textbf{c}. Starting from an initial density-wave state, the random modulation of $\hat{U}_{\pm}$ destabilizes the system and hence induces heating. The heating rate can be significantly suppressed by imposing a dipolar structure into the random driving sequence, such that the elementary building blocks now read $\hat{U}_{+}\hat{U}_{-}$ and $\hat{U}_{-}\hat{U}_{+}$. Similarly, the $n$-th  multipole can be recursively constructed by anti-aligning two $(n-1)$-th order operators, and in the $n{\to}\infty$ limit it converges to the self-similar Thue-Morse driving~\cite{Nandy2017,Mori2021,pilatowsky2025critically}.  
    
    We first benchmark our experiments on 8 qubits, then we gradually enlarge the system size up to 78 qubits. To quantify the heating process, we experimentally monitor the decay of the particle imbalance. In addition, we measure the growth of the entanglement entropy, thus going beyond established experimental results on driven systems where typically only the evolution of local observables is accessible~\cite{RubioAbadal2020, Peng2021,Beatrez2023,He2023}. 
    We perform more than 1,000 driving cycles, and such a long time scale allows us to capture the long-lived prethermal plateau in the high-frequency regime. Moreover, we experimentally verify the crucial role of the temporal multipolar correlation in stabilizing the system: The heating rates follow a power-law dependence on the driving frequency, with a universal scaling exponent of approximately $2n {+} 1$, in accordance with the original theoretical analyses of the heating processes active for RMD~\cite{ZhaoRandom2021,Mori2021}. 
    
    Then by selecting different subsystem configurations, we demonstrate a non-uniform spatial entanglement distribution and observe the crossover from the area-law to the volume-law entanglement scaling within the prethermal regime. The onset of heating further accelerates entanglement growth, posing a significant challenge to account for it with numerical simulations within a realistically accessible computational time~\cite{FloquetPrethermalization2022}. Advanced tensor network numerical techniques, such as grouped matrix product states (GMPS) and projected entangled pair states (PEPS), struggle to keep pace with the rapid entanglement growth. Therefore, our experiment potentially demonstrates the quantum advantage in emulating the entire heating dynamics towards the maximally entangled infinite-temperature state~\cite{SimulatingPrethermalization2023}.

    \begin{figure*}[t]
	\centering
	\includegraphics[width=0.97\linewidth]{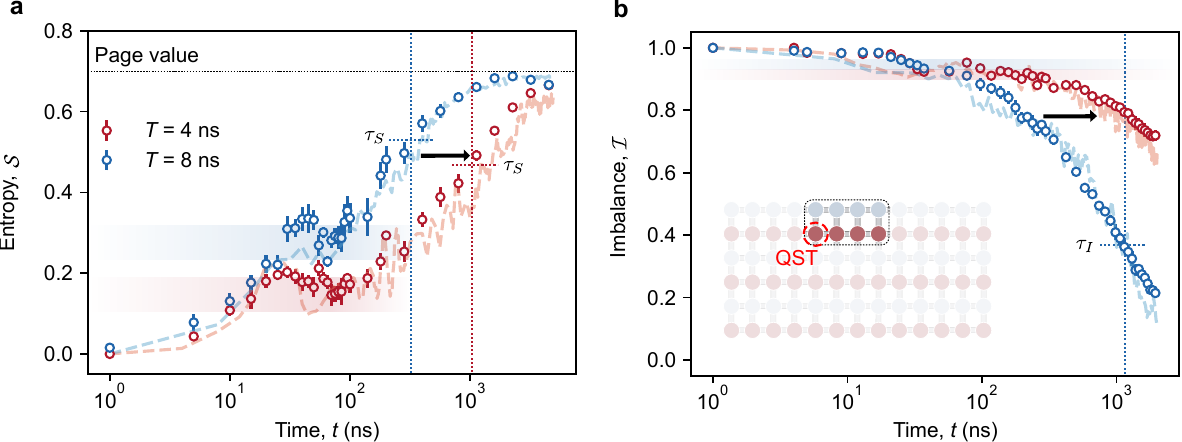}
	\caption{\textbf{Prethermalization by random multipolar driving (RMD) in an $8$-qubit system.} \textbf{a} (\textbf{b}), Experimental data showing the dynamics of the subsystem entanglement entropy (imbalance), for driving periods $T=4$~ns and 8~ns, where the prethermal plateau is clearly probed. Both the lifetime of the entanglement plateau, $\tau_S$, and the imbalance decay time scale, $\tau_I$, can be prolonged by increasing the driving frequency. 
	We implement the $1$-RMD protocol, and the system comprises 8 qubits $\{Q_{x, y}\}$ (the inset in \textbf{b}), indexed by $x=\{5, 6, 7, 8\}$ and $y=\{1, 2\}$. 
	We use quantum state tomography (QST) to determine the von-Neumann entropy of the qubit $\{Q_{5, 2}\}$. Error bars indicate 1SD of the experimental results, averaged over 10 independent RMD sequences. The light dashed curves depict numerical results for benchmarking.}
	\label{fig2}
    \end{figure*}
    
	~\\
	\textbf{Experimental setup}
	~\\
	Our experiments are performed on a flip-chip superconducting processor arranged, in a $6 \times 13$ square lattice array (Fig.~\ref{fig1}\textbf{a}), with 78 qubits and 137 couplers between all neighbor qubits. The qubits are labeled as $Q_{x, y}$, with $x$ ranging from $1$ to $13$ and $y$ from $1$ to $6$, see Fig.~\ref{fig1}\textbf{d}. Since the anharmonicity is designed to be much larger than the hopping strength, with an average value of $-$200~MHz, the system can be described as the non-integrable 2D hard-core Bose-Hubbard model~\cite{yan_strongly_2019, Deng2024}. The effective Hamiltonian reads ($\hbar=1$):
	\begin{equation}
		\hat{H}=\sum_{j}\omega_j\hat{n}_j+\sum_{\langle i, j\rangle}g_{i,j}\left(\hat{a}^{\dagger}_i\hat{a}_j+\mathrm{H.c.}\right),
	\end{equation}
	where $i, j$ labels the qubit in the group $\{Q_{x,y}\}$, $\hat{n}_j$ denotes particle number operator at the $j$-th site, $\hat{a}_j^{\dagger}$($\hat{a}_j$) is the creation (annihilation) operator, $\omega_j$ is the on-site potential, and $g_{i, j}$ is the hopping strength between two nearest-neighbor qubits. Device information can be found in (SM)~\cite{supp_cite}. Owing to the significant progress in coherence time, scalability, and controllability of superconducting quantum circuits~\cite{yan_tunable_2018,Krantz2019, Liang2023}, this platform has emerged as a powerful system for exploring complex quantum phenomena that requires precise manipulation.

	Our target elementary operators in the $n$-RMD protocol are generated as $\hat{U}_{\pm}{=}\exp(-\mathrm{i} \hat{H}_{\pm} T)$, where $T$ denotes the characteristic timescale (noted as the driving period below). In our experiments, 
	\begin{equation}
	\label{eq.drive}
		\hat{H}_{\pm}=\hat{H}_{\kappa}+(1\pm\delta h)\hat{H}_p,
	\end{equation}
	with a dimensionless parameter $\delta h$ characterizing the driving amplitude. Here, the on-site term exhibits a uniform potential along the $x$-axis and a staggered arrangement along the $y$-axis
	\begin{equation}
		\hat{H}_p=h_0\sum_{x,y}[1-(-1)^y]\hat{a}^{\dagger}_{x,y}\hat{a}_{x,y},
		\label{eq.staggertedpotential}
	\end{equation}
	with the hopping term isotropic in both spatial directions reads
	\begin{equation}
	\hat{H}_{\kappa}=J\sum_{x,y}\left(\hat{a}^{\dagger}_{x,y}\hat{a}_{x+1,y}+\hat{a}^{\dagger}_{x,y}\hat{a}_{x,y+1}+\mathrm{H.c.}\right).
% 		\hat{H}_{\kappa}=J\sum_{x,y}\left(\hat{a}^{\dagger}_{x,y}\hat{a}_{x+1,y}+\mathrm{H.c.}\right)+J\sum_{x,y}\left(\hat{a}^{\dagger}_{x,y}\hat{a}_{x,y+1}+\mathrm{H.c.}\right).
	\end{equation}
	The parameters are chosen as $J/2\pi=2$~MHz, $\delta h=1.2$, and $h_0/2\pi=10$~MHz; and $T$ ranges from 3~ns to 8~ns. The average decoherence time ($T_1$) of our device is 26.4~$\mu$s~\cite{supp_cite}, allowing us to experimentally implement more than 1,000 driving cycles, before any notable decoherence occurs. 
	
	For $1$-RMD, the protocol involves a random sequence of two dipolar operators $\hat{U}_{+}\hat{U}_{-}$ and $\hat{U}_{-}\hat{U}_{+}$; and for $2-$RMD, the elementary building blocks become $\hat{U}_{+}\hat{U}_{-}\hat{U}_{-}\hat{U}_{+}$ and $\hat{U}_{-}\hat{U}_{+}\hat{U}_{+}\hat{U}_{-}$. 
	Experimentally, implementing the driving protocol requires precise temporal modulation of the pulse signal, especially in the high-frequency regime ($T{\approx}3$~ns): Qubits with odd $y$-index are biased to the working point, while the pulse signal following an RMD sequence is implemented on qubits with even $y$-index. Through careful calibration of the Z pulse distortion and crosstalk, combined with precise timing alignment via Floquet engineering, we achieve highly accurate RMD control, see details SM~\cite{supp_cite}. 
% 	The average energy relaxation time is about 26.4~$\mu$s, much longer than the experimental time $\sim 2$~$\mu$s.
% 	\hz{we should compare the $T_1, T_2$ with the allowed number of driving cycles here, to explicitly state that coherence time is much longer than the experimental time ($\sim 2\mu$s).}\yuliu{The dephasing time $T_2^*$ is about 1.2~$\mu$s. However, we believe the real dephasing time during the dynamics is longer than $T_2^*$. We can argue it by citing some papers if any reviewers question this.}
	
	As shown in Fig.~\ref{fig1}\textbf{d}, we initialize the system as a density wave ordered product state $\ket{\psi_0}$, where all lattice sites with even $y$-index are occupied, resulting in the maximal Hilbert space dimension. In the high-frequency limit, $T{\to} 0$, the early time evolution of the system can be described by an effective Hamiltonian, $\hat{H}_{\mathrm{eff}}{=}(\hat{H}_+{+}\hat{H}_-)/2$, where the staggered potential along the $y$-axis, Eq.~\ref{eq.staggertedpotential}, stabilizes the initial density wave order. However, for any finite $T$, switching between $H_+$ and $H_-$ destabilizes the system, and eventually, it heats up to the infinite-temperature state. Therefore, the particle number imbalance serves as 
	a good diagnostics of this heating process 
	\begin{equation}
	    \mathcal{I}=\frac{1}{N_0}\sum_{j\in\{x,y\}}\bra{\psi_0}\hat{Z}_j(t)\hat{Z}_j(0)\ket{\psi_0},
	\end{equation}
	with $\hat{Z}_j=2\hat{n}_j-1$, and the initial particle number $N_0$. In addition to these local observables, we also study the time evolution of the entanglement entropy, $S{=}{-}\textrm{Tr}[\rho\ln\rho]$ where $\rho$ denotes the reduced density matrix of a subsystem, to capture the growth of non-local correlations between a subsystem and its complement~\cite{Abanin2019}. It provides valuable information for estimating the numerical complexity in simulating the many-body dynamics~\cite{Osterloh2002, Eisert2010, Khemani2017, karamlou_probing_2024}. In the experiment, we perform quantum state tomography (QST)~\cite{Xu2018} on a subsystem to reconstruct $\rho$ at different stages of the heating process. 

	\begin{figure*}[t]
	\centering
	\includegraphics[width=0.97\linewidth]{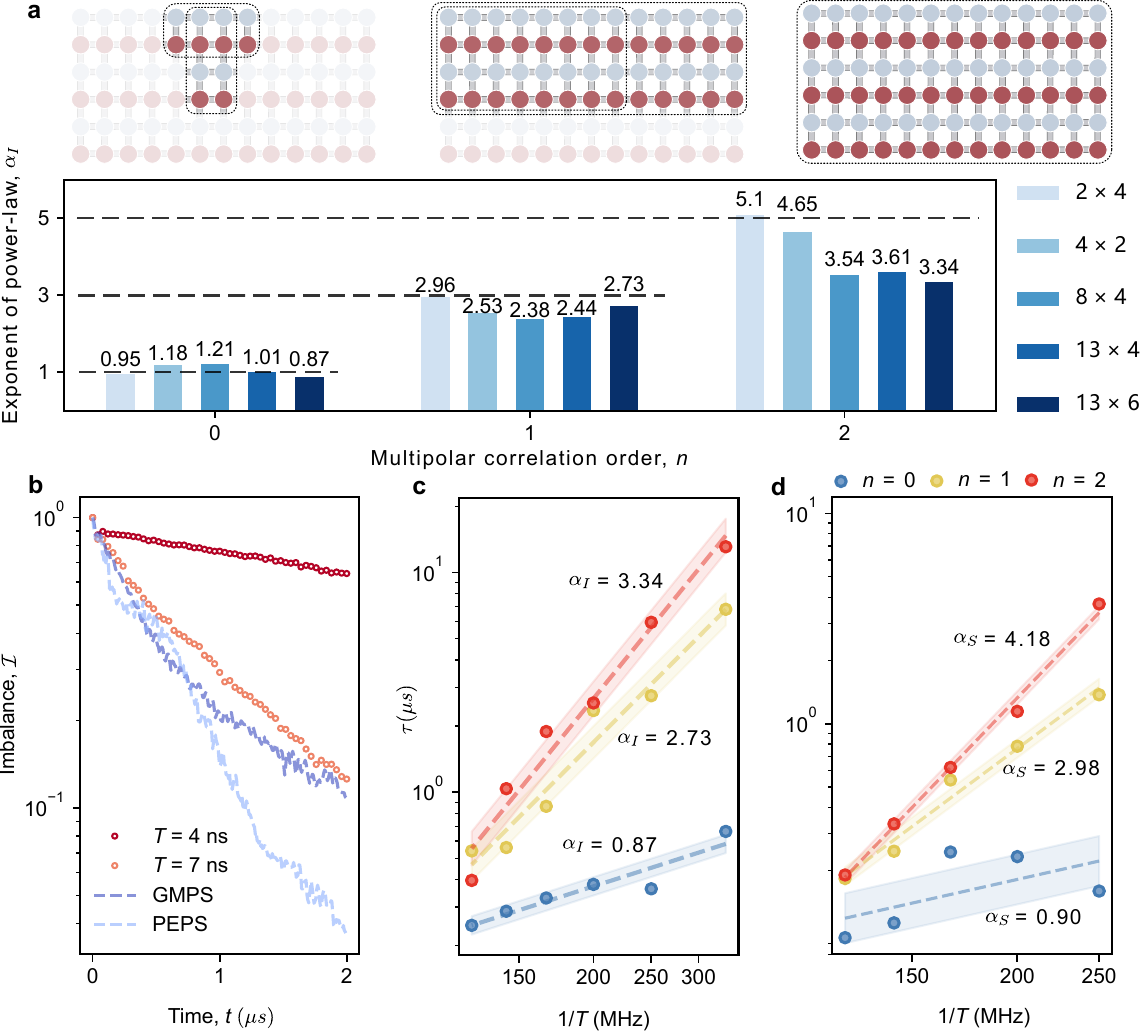}
	\caption{\textbf{Scaling behavior of the prethermal lifetime with driving frequency.} \textbf{a}, Various system configurations and their corresponding power-law scaling exponents. These scaling exponents closely follow the theoretically predicted value of $\alpha(n){=}2n{+}1$ (the dashed lines). \textbf{b}, Decay of the imbalance shown in a log scale. The markers represent the experimental data, while the dashed curves are obtained numerically using the grouped matrix product states (GMPS) with bond dimension $\chi{=}96$ and projected entangled pair states (PEPS), for $T{=}7$~ns and $4$~ns, respectively, with 2-RMD drives. Numerical methods can only capture the early time evolution. \textbf{c} (\textbf{d}), The power-law scaling of the prethermal lifetime obtained from the dynamics of the imbalance $\tau_I$ (the subsystem entanglement entropy $\tau_S$) in the 78-qubit system. A larger multipolar order $n$ significantly suppresses the heating rates. The subsystem for calculating the entanglement entropy comprises qubits $\{Q_{5, 3}, Q_{6, 3}\}$. }
	\label{fig3}
	\end{figure*}
	
	~\\
	\textbf{Characteristics of prethermalization by RMD}
	~\\
    We first benchmark our experiments on 8 qubits, which can be efficiently simulated numerically. Then, we gradually increase the system size, as shown in Fig.~\ref{fig3}\textbf{a}, and finally use all 78 qubits, allowing for a detailed analysis of heating dynamics and the entanglement entropy growth.

     Figure~\ref{fig2}\textbf{a} displays the measured entanglement entropy dynamics of a single qubit under 1-RMD, in a small-scale system of 8 qubits and 10 couplers, for $T{=}4$~ns and $T{=}8$~ns respectively. The entropy evolution exhibits several distinct features: (1) Starting from $S{=}0$, as expected for a product initial state, the entanglement entropy rapidly increases within a short time; (2) A distinct prethermal plateau appears within the time interval $20{\lesssim} t {\lesssim} 100$~ns, hallmarking the notable suppression of energy absorption. This plateau can be described by a quasi-conserved effective Hamiltonian, which can be perturbatively constructed using a generalized Floquet-Magnus expansion~\cite{supp_cite,Mori2021}; (3) At later times, $S$ deviates from the plateau and continues increasing. Around $t{\gtrsim} 2,000$~ns, it saturates at $S_M{\approx} 0.7$, in accordance with the Page value, $S_P\approx L\ln 2$~\cite{Page1993}, where $L$ denotes the subsystem size, indicating that the system completely thermalizes. We also perform exact numerical simulations (light dashed lines in Fig.~\ref{fig2}\textbf{a}), which match well with our experimental observations. The lifetime of the prethermal plateau,  $\tau_S$, can be quantitatively determined by numerically fitting the entanglement entropy growth to the ansatz $S{\sim} S_M(1{-}e^{-t/\tau_S})$~\cite{Machado2020} in the intermediate region $100{\lesssim} t{\lesssim} 2,000$~ns. By reducing the driving period from $T{=}8$~ns to $T{=}4$~ns, $\tau_S$ significantly increases from $0.25$~$\mu$s to $1.03$~$\mu$s, indicating a strong suppression of the heating rate. 
	
	\begin{figure*}[t]
	\centering
	\includegraphics[width=0.95\linewidth]{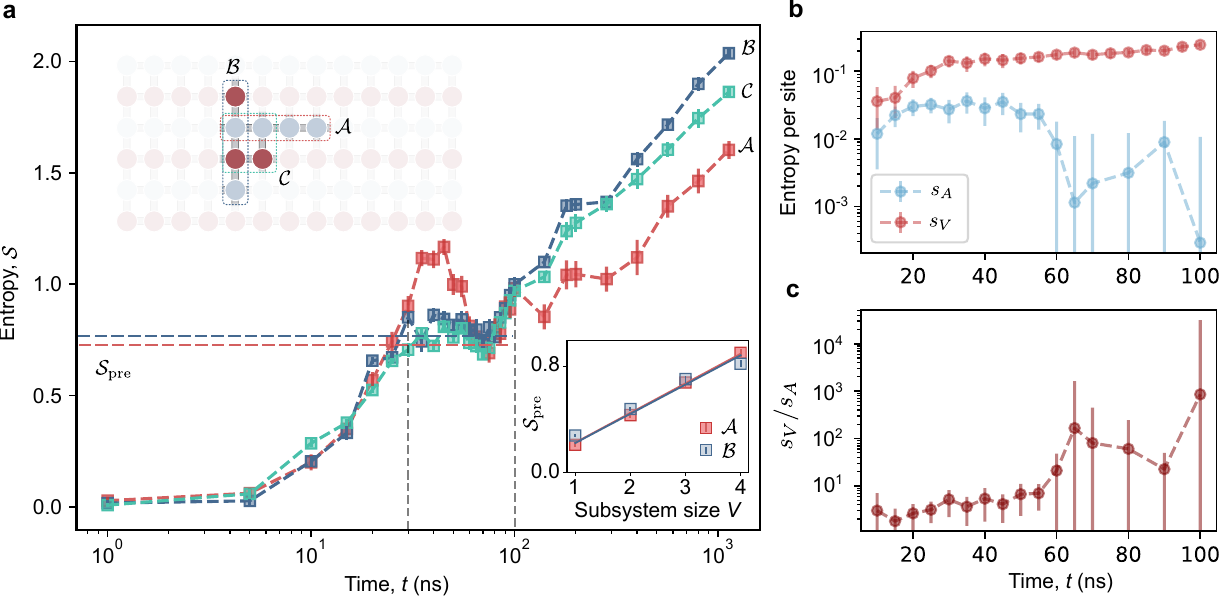}
	\caption{\textbf{Entanglement dynamics and volume law scaling.} \textbf{a}, Dynamics of the entanglement entropy for different subsystem configurations, involving 4 qubits in the 78-qubit system. All subsystems dynamics enter a long-lived prethermal regime, among which $\mathcal{A}$ exhibits pronounced oscillatory dynamics. The inset depicts the entanglement entropy averaged over the prethermal regime ($30~ \text{ns}{\leq} t{\leq} 100~\text{ns}$, denoted between two black dashed vertical lines in panel \textbf{a}), as a function of volume $V$ for subsystems $\mathcal{A}$ and $\mathcal{B}$. \textbf{b}, The volume ($s_V$) and the area entanglement entropy ($s_A$) per site, numerically fitted by analyzing subsystems of varying volumes and areas at different times. The error bars correspond to 1SD of the fitting parameter. \textbf{c}, Ratio $s_V/s_A$ at different times, as a quantitative measure to distinguish between the area-law and the volume-law scaling. As time evolves, a clear crossover from the area-law to volume-law scaling is observed. Here, we use $1$-RMD protocol and $T{=}4$~ns to perform the experiment. }
	\label{fig4}
	\end{figure*}
	
	A similar feature also appears during the decay of the imbalance.
	 As shown in Fig.~\ref{fig2}\textbf{b}, we observe that, starting from the initial value $\mathcal{I}{=}1$, the imbalance decays towards $0$ at later times, and a high-frequency drive significantly slows down this decay. 
	 The characteristic decay time scale $\tau_I$ can be extracted by fitting the imbalance evolution to an exponentially decaying function $\mathcal{I}\sim e^{-t/\tau_I}$. For $T=8$~ns and $T=4$~ns, the time scales are $1.2$~$\mu$s and $7.4$~$\mu$s, respectively.
	
	~\\
	\textbf{\textbf{Tunable heating rate by RMD}}
	~\\
	To further study the suppression of the heating rate, we perform a similar RMD protocol and 
	examine the dependence of the prethermal lifetime on the driving frequency for a different multipolar order $n$.  Experimentally, we implement the driving protocol on larger systems with distinct geometries, as shown in Fig.~\ref{fig3}\textbf{a}. In particular, in Fig.~\ref{fig3}\textbf{b}, we present the time evolution of the imbalance for the 78-qubit system driven by the 2-RMD protocol using $T{=}4$~ns and $7$~ns and (orange and red dots). In both cases, the imbalance decays exponentially, in agreement with both numerical~\cite{Machado2020} and experimental observations during the heating process~\cite{RubioAbadal2020}.  
	In Fig.~\ref{fig3}\textbf{c} and \ref{fig3}\textbf{d}, we use a log-log scale and show the prethermal lifetimes $\tau_I$ and $\tau_S$ for different $1/T$. Data points approximately follow a straight line, suggesting a power-law decay, $\tau_{I, S}{\sim} (1/T)^{\alpha}$. 
	This power-law behavior is in sharp contrast to conventional Floquet systems' with local interactions, where the prethermal lifetime scales exponentially with the increase the driving frequency $\tau_{I, S}{\sim} e^{1/T}$~\cite{Peng2021, He2023}. Note that the scaling exponent $\alpha$ is now tunable and exhibits a strong dependence on the multipolar order in the high-frequency regime. This is a significant feature of RMD systems, which cannot be achieved by either Floquet or other quasi-periodic drives. The reason is that the multipolar structure suppresses the low-frequency components of the drive, thereby constraining heating. A generalized Floquet-Magnus theory and a Fermi’s golden rule argument predicts the relation $\alpha{=}2n{+}1$ ~\cite{ZhaoRandom2021, Mori2021}. 
	
	Experimentally, for the 78-qubit system and $n{=}\{0, 1, 2\}$, we numerically fit the exponent based on the prethermal lifetime of the entanglement entropy plateau and obtain $\alpha_S=\{0.904, 2.976, 4.182\}$.
	Based on the imbalance decay, we obtain the exponents  $\alpha_I{=}\{0.871, 2.727, 3.344\}$. In Fig.~\ref{fig3}\textbf{a}, we summarize the scaling exponents $\alpha_I$, which reveal a positive correlation between $\alpha_I$ and $n$ across all investigated system sizes and configurations. More experimental data on the corresponding time evolutions can be found in~\cite{supp_cite}. In particular, the scaling exponents for $n=0$ and $1$ closely follow the theoretical prediction (black dashed horizontal lines in Fig.~\ref{fig3}\textbf{a}). For $n=2$, it is particularly challenging to obtain a converging scaling exponent, which requires an extremely fast drive and long evolution time, see detailed discussions in SM~\cite{supp_cite}. Here, we implement approximately 1,000 driving cycles to clearly distinguish the scaling behaviors for $n=1$ and $2$. This also highlights the exceptional long-time stability and controllability of \textit{Chuang-tzu 2.0}, which allows us to precisely engineer the temporal correlations and manipulate the quantum thermalization process.
	
	To validate our experimental observations, we employ advanced numerical techniques based on tensor network representations~\cite{Cirac2021}
	%, specifically the projected entangled pair states (PEPS) and the grouped matrix product staets (BMPS), 
	to approximately simulate the many-body dynamics. We initially developed the projected entangled pair states (PEPS) with a single update for simulations. PEPS is a 2D tensor network and is feasible for our circuit system. However, due to the lack of canonical form, PEPS is not accurate in long-time simulations. The simulation results are shown in Fig.~\ref{fig3}\textbf{b}, where we can see that shortly after a few driving cycles ($t\approx 0.1$~$\mu$s), the PEPS simulations exhibit significant deviations from the experimental data. To improve the numerical accuracy, we further developed the grouped MPS (GMPS) method.  Compared to the conventional matrix product states (MPS) method, the GMPS method merges certain tensors in the conventional MPS into a giant tensor, thereby avoiding truncations inside the giant tensor and the swapping operations in MPS. This strategy is particularly efficient in simulating quasi-1D systems such as our system. As shown in Fig.~\ref{fig3}\textbf{b}, GMPS correctly reproduces the experimental data at early times, $t\lesssim 0.3$~$\mu$s, and later deviations from the experimental data also become visible. {In SM~\cite{supp_cite} , we systematically study the performance of GMPS and observe that its estimated fidelity significantly decays as heating occurs (fidelity drops below 0.05 when $t{\approx} 1.5$~$\mu$s). The rapid growth of entanglement during this process limits both its accuracy and efficiency. Therefore, our quantum processor allows us to simulate the entire heating dynamics towards the highly-entangled infinite temperature states that is extremely costly to simulate classically ~\cite{Arute2019,Zhong2020,Wu2021,Daley2022, Gao2025, King2025}.}
	
	~\\
	\textbf{\textbf{Entanglement dynamics and scaling}}
	~\\
	We further investigate the subsystem size dependence of the entanglement entropy. Specifically, we perform QST on various subsystems, each comprising up to $4$ sites from the 78-qubit system. As shown in Fig.~\ref{fig4}\textbf{a}, these configurations exhibit distinct spatial arrangements: the subsystem $\mathcal{A}$ is aligned with the $x$-axis, $\mathcal{B}$ is oriented along the $y$-axis, and $\mathcal{C}$ forms a $2\times 2$ square lattice. Figure~\ref{fig4}\textbf{a} depicts the entanglement entropy dynamics for different subsystems, all exhibiting the long-lived prethermal regime. Moreover, the entanglement dynamics strongly depend on the specific choice of subsystems:  $\mathcal{A}$ exhibits pronounced oscillatory dynamics during the prethermal regime, while $\mathcal{B}$ and $\mathcal{C}$ quickly saturates at a plateau, suggesting a non-uniform entanglement distribution in 2D. These oscillations originate from the coherent particle exchange between qubits with even and odd $y$-index, stabilized by the staggered potential in the effective Hamiltonian $\hat{H}_{\mathrm{eff}}$, and can also be validated via GMPS simulations, see~\cite{supp_cite}. 
% 	This 
% 	exhibiting anisotropic entanglement dynamics in the early-time evolution. This behavior can be understood through the effective single-particle Hamiltonian $\hat{H}'=J\sigma_x+\Delta \sigma_z$, where $\sigma_x, \sigma_z$ are Pauli matrices, $\Delta$ represents the potential barrier difference between the subsystem and and its complement, governing the underlying dynamics. In specific, subsystem $\mathcal{A}$ exhibits the largest average value of $\Delta$, resulting in the most pronounced oscillation patterns in its entanglement dynamics.
	
	To quantify the entanglement entropy scaling, we calculate the averaged entropy, $\mathcal{S}_\textrm{pre}$, in the prethermal regime to reduce the temporal fluctuations. In the inset of Fig.~\ref{fig4}\textbf{a}, we show $\mathcal{S}_\textrm{pre}$ as a function of the linear subsystem size $V$ for $\mathcal{A}$ (red) and $\mathcal{B}$ (blue), and linear dependence can be observed. {For other possible subsystem configurations, in practice, one can approximately distinguish different contributions using the ansatz~\cite{karamlou_probing_2024}
	\begin{equation}
	    \mathcal{S}_\textrm{pre}(\rho_X)=s_AA_X+s_VV_X,
	\end{equation}
	where $\rho_X$ denotes the reduced density matrix of a subsystem $X$, $A_X$ and $V_X$ correspond to  the subsystem's area and volume, respectively. The ratio $s_V/s_A$ quantifies the degree to which the state adheres to area-law or volume-law entanglement scaling. By analyzing 12 non-repetitive subsystems by varying volumes and areas, we numerically fit $s_V$ and $s_A$ and show their time evolutions in Fig.~\ref{fig4}\textbf{b} and the ratio $s_V/s_A$ in Fig.~\ref{fig4}\textbf{c}. At early times ($t{\leq}30$ ns), the system stays close to the density wave ordered product state, and $s_V/s_A$ remains small. A notable increase occurs around $t{\approx}50$ ns, and this ratio becomes substantially larger than 1, suggesting a crossover from area-law to volume-law scaling.
     At later times, entanglement increases and still obeys the volume-law scaling, so the system's dynamics become further computationally intractable}.

	~\\
	\textbf{Conclusion and outlook}
	~\\
% 	Although our experiment is performed on a superconducting processor, the underlying heating control mechanism is readily applicable to different quantum simulator platforms. 
   We present a systematic experimental study of the non-equilibrium dynamics of 2D interacting systems driven by the $n$-RMD protocol on a 78-qubit superconducting processor. Its precise and stable pulse sequence allows for the demonstration of the long-lived prethermal plateau and suppressed heating rates, which exhibit the universal algebraic scaling $T^{2n+1}$ in the high-frequency regime. The possibility of driving closed quantum systems with temporal randomness while avoiding heating paves the way for engineering novel non-equilibrium phases of matter beyond the conventional Floquet paradigm.
   
   By performing QST, we observe the growth of the subsystem entanglement entropy during the entire heating process. Although the system eventually evolves towards the featureless infinite-temperature state, during the prethermal regime, we observe a non-uniform spatial entanglement distribution. In particular, subsystem $\mathcal{A}$ exhibits coherent oscillatory entanglement dynamics. These observations shed light on the understanding of the microscopic generation of entanglement, especially in higher dimensions, where the subsystem configurations can exhibit a rich geometric structure. 
   
Furthermore, by analyzing the entanglement entropy for various subsystems, we demonstrates the crossover from area-law to volume-law entanglement scaling as time evolves. Such rapid entanglement growth poses significant challenges in simulating our experiments using classical algorithms. Therefore, our work demonstrates a potential quantum advantage in emulating the entire non-equilibrium heating dynamics in driven systems. 

It will be interesting to further explore the initial state dependence and the spatial non-homogeneity in the heating process, as well as the stability of MBL and anomalous topological phases~\cite{Anomalous2022} in randomly driven systems. In addition, identifying different prethermalization mechanisms for much broader classes of non-periodic drives remains an open and interesting direction. Finally, although our experiment is performed on a superconducting processor, the underlying heating control mechanism is readily applicable to different quantum simulator platforms and can stabilize sought-after non-equilibrium phenomena in driven systems. 

%   Its heating rate can be precisely controlled through the engineered pulse sequence with exceptional tunability, as quantitatively demonstrated by measurements of both the imbalance and subsystem entanglement entropy. 
%   By resolving thermalization dynamics in a large-scale system up to 78 qubits, which is highly challenging for the state-of-the-art classical computers to exactly simulate, our work highlights the remarkable capabilities of analogue-digital quantum simulators~\cite{Andersen2025}. Looking ahead, our findings open the door to several promising directions. First, it is interesting to study heating dynamics with systematically prepared initial states, providing crucial insights into state-dependent thermalization behavior. Second, exploring the anisotropic entanglement growth in higher-dimensional systems could reveal new universal features of quantum thermalization. Finally, elucidating the microscopic mechanisms underlying thermalization processes would complement our current understanding of the system's statistical properties, offering a more complete picture of non-equilibrium quantum dynamics.

	\bibliography{main.bib}
	
	~\\
	\textbf{Acknowledgments}
	~\\
	P.Z. and K.C. thank Bozhen Zhou for helpful discussions on the MPS simulations. We thank Takashi Mori for his previous theoretical contributions.
	We thank the support from the Synergetic Extreme Condition User Facility (SECUF) in Huairou District, Beijing. Devices were made at the Nanofabrication Facilities at Institute of Physics, CAS in Beijing. This work was supported by National Natural Science Foundation of China (Grants Nos.~T2121001, T2322030, 92265207, 12122504, 12247168, 12325501, 12047503, 12247104, 12474214, 12475017, 92365301), Innovation Program for Quantum Science and Technology (Grant No.~2021ZD0301800), Beijing Nova Program (No.~20220484121), Natural Science Foundation of Guangdong Province (Grant No.~2024A1515010398).
    
    ~\\
	\textbf{\textbf{Competing interests}}
	~\\
	The authors declare no competing interests.

% 	with the characteristic timescale $T=4$~ns, 

% \textbf{b}, The prethermal lifetime obtained from the dynamics of the imbalance $\tau_i$, as a fuction of inverse characteristic timescale $1/T$.
	
% 	\textbf{c}, The prethermal lifetime obtained from the dynamics of entanglement $\tau_s$, versus $1/T$. Both definitions of the lifetime follow the same power-law scaling $(1/T)^{\alpha}$, with $\alpha\approx 2n+1$.
	\vspace{-1mm}
\clearpage
\onecolumngrid

\newcommand{\beginsupplement}{%
	\setcounter{table}{0}
	\setcounter{figure}{0}
	\setcounter{page}{1}
	\renewcommand{\thesection}{\:S\arabic{section}}
	\setcounter{section}{0}
	\setcounter{equation}{0}
	\renewcommand{\theequation}{\arabic{equation}}
}

	\begin{center}
		\textbf{\large{\textit{Supplementary Materials for:} \\ \smallskip Prethermalization by Random Multipolar Driving on a 78-Qubit Superconducting Processor}} \\\smallskip
	\end{center}

% % 	%\preprint{APS/123-QED}
% 	\title{Supplementary Materials for:\\ ``Prethermalization by Random Multipolar Driving on a 78-Qubit Superconducting Processor''}% Force line breaks with \\

% 	\date{\today}% It is always \today, today,
	%  but any date may be explicitly specified
	%\keywords{Suggested keywords}%Use showkeys class option if keyword
	%display desired
% 	\twocolumngrid
	
	\beginsupplement
	\tableofcontents
	
	\renewcommand{\figurename}{Supplementary Fig.}
	\renewcommand{\tablename}{Supplementary Table}
	
	%\tableofcontents
    
     \section{Prethermalization and numerical simulation}
	 \subsection{Prethermalization in Floquet systems}
	 Periodically driven (Floquet) quantum systems can
    engineer novel far-from-equilibrium phases of matter absent in static systems~\cite{Eckardt2017}. These system can be analysed within the framework of Floquet theory. Consider the Hamiltonian $\hat{H}$ with temporal periodicity, $\hat{H}(t)=\hat{H}(t+T)$, the Floquet theorem states that the time evolution operator can be written as
    \begin{equation}
        \hat{U}(t)=\hat{P}(t)\mathrm{e}^{-\mathrm{i}\hat{H}_Ft},
    \end{equation}
    where $\hat{H}_F$ is the time-independent Floquet Hamiltonian, and $\hat{P}(t)$ is the stroboscopic operator with period $T$. In the high frequency regime, the dynamics can be approximated by the Hamiltonian $\hat{H}_F$, known as the effective Hamiltonian. In general, an exact closed-form expression for $\hat{H}_F$ is difficult to obtain, and a common scheme is to use the Floquet-Magnus expansion (FME), which is a perturbative expansion in the driving frequency
    \begin{equation}
        \hat{H}_F=\sum_{n=0}^{\infty}\frac{1}{\Omega^n}\hat{H}_F^{(n)},
    \end{equation}
    with the driving frequency $\Omega=2\pi/T$. The lowest-order effective Hamiltonian is the time average of system Hamiltonian $\hat{H}$ over a period $T$
    \begin{equation}
        \hat{H}_F^{(0)}=\frac{1}{T}\int_{0}^T\!\mathrm{d}t\;\hat{H}(t).
    \end{equation}
    
	Unlike static systems that conserve energy, Floquet systems typically absorb energy from the drive~\cite{Ueda2020, DAlessio2014}. The nature of equilibrium states under such heating becomes clearer through the analysis of FME convergence. When the FME converges, the effective Hamiltonian $\hat{H}_F$ can be effectively represented by a summation of local operators. In this case, the system is expected to thermalize to a steady state that, at the local level, becomes effectively indistinguishable from the microcanonical ensemble of a static system described by $\hat{H}_F$. However, it is widely believed that the FME diverges in generic many-body interacting systems, indicating the absence of a quais-local $\hat{H}_F$. For a non-integrable Floquet system, heating leads to the eventual thermalization where the system reaches the featureless infinite-temperature state.
	 
    Theoretical studies on the long-time behaviors of Floquet systems reveal that, the FME may converge to the time-averaged Hamiltonian when driving periods are sufficiently short, such that the driving frequency $\Omega$ is much larger than the local energy scale of many-body system. In this regime, Floquet prethermalization appears before the onset of noticeable heating, giving rise to an exponentially long-lived prethermal regime governed by the prethermal Hamiltonian
    \begin{equation}
        \hat{H}_{\textrm{pre}}=\sum_{n=0}^N\frac{1}{\Omega^n}\hat{H}_F^{(n)},
    \end{equation}
  obtained by truncating FME at the order $N$. In contrast to using many-body localization (MBL) to suppress heating~\cite{Abanin2019}, Floquet prethermalization is generally applicable to clean systems without the need of strong spatial disorder.
    
    The exponential suppression of the heating rate can be justified by noticing the residual part of the Floquet Hamiltonian satisfying $\|\hat{H}_F-\hat{H}_{\textrm{pre}}\|<\mathrm{e}^{-\mathcal{O}(1/\Omega)}$, with $\|\cdot\|$ denoting the norm of an operator. Higher-order processes lead to exponentially small corrections that become noticeable after an exponentially long time scales~\cite{RigorousBound2016}.     Alternatively, one can also use a simple linear response theory--the Fermi's golden rule (FGR)~\cite{ExponentiallySlowHeating2015,ZhaoRandom2021}--to understand this exponential suppression. We start with a generic Floquet system
    \begin{equation}
        \hat{H}=\hat{H}_0+g(t)\hat{K},
    \end{equation}
    where $g(t)$ is taken the Fourier series form $g(t)=\sum\limits_{m>0}g_m\sin (m\Omega t)$, $\hat{H}_0$ and $\hat{K}$ are time-independent operators. For Fourier mode $m$, the heating rate is given by
    \begin{equation}
        \Gamma_m(t)=\frac{\dot{E}_m(t)}{E_{\infty}-E(t)},
    \end{equation}
    where $E_{\infty}$ denotes the energy at infinite temperature with $E_{\infty}=0$ for our model. The transition rate of the mode $\dot{E}_m(t)$ can be obtained from the FGR
	\begin{equation}
	    \dot{E}_m(t) = 2\pi g_m^2 \sum_{i,f} | \langle E_f^0 | \hat{K} | E_i^0 \rangle|^2 ( E_f^0 - E_i^0) P_i^0(t)\cdot \delta( E_f^0 - E_i^0 \pm m\Omega),
	\end{equation}
	with $E_i^0, E_f^0$ being eigenstates of $\hat{H}_0$, and $P_i^0(t)=\bra{E_i^0}\rho(t)\ket{E_i^0}$.
    Combining all Fourier modes, the heating rate of the system is
	\begin{equation}
	    \Gamma(t)=\sum_{m>0}\Gamma_m(t),
	\end{equation}
	with the $m$-mode rate
	 \begin{equation}
	 \label{eq.Floquetrate}
	     \Gamma_m = g_m^2 A e^{-m\Omega/\epsilon},
	 \end{equation}
	 where $A$ and $\epsilon$ are $\Omega$-independent constants.
	 
	 \subsection{Prethermalization by random multipolar driving}
	 Non-periodically driven systems have attracted notable research interest during recent years, which go beyond the constraint imposed by the strict temporal periodicity. We focus on a family of random but structured driving protocol, random multipolar driving (RMD), characterized by the multipolar correlation $n$. 
	 This protocol involves two elementary evolution operators $\hat{U}_+$ and $\hat{U}_-$, generated by two Hamiltonians $\hat{H}_{\pm}$. A random selection of $U_{\pm}$ leads to $0-$RMD sequence and normally this quickly destabilizes the system and induces heating. The heating rate can be significantly suppressed by imposing a dipolar structure into the random driving sequence, such that the elementary building blocks now read $\hat{U}_{+}\hat{U}_{-}$ and $\hat{U}_{-}\hat{U}_{+}$. $1-$RMD protocol then contains a random sequence of these two building blocks. Similarly, the $n$-th  multipole can be recursively constructed by anti-aligning two $(n-1)$-th order operators, and in the $n{\to}\infty$ limit it converges to the self-similar Thue-Morse driving~\cite{Mori2021}.  A detailed discussion on the properties of RMD sequences is presented in \cite{ZhaoRandom2021} and here we elaborate on how to estimate the heating rate.
	 
	 A purely random sequence exhibits a flat and continuous distribution of different frequencies in the Fourier space. Similarly, $n$-RMD sequence also exhibits a continuous spectrum. However, the overall envelope of the distribution is modified to a power law $x^n$~\cite{ZhaoRandom2021}, due to the presence of the multipolar correlation in time. Therefore, to estimate the heating rate one can first approximate the time-dependent RMD signal as
	 \begin{equation}
	     g(\tau) \hat{K} = \int\mathrm{d}x g_x\sin(x\Omega\tau)\hat{K},
	 \end{equation}
    where $g_x = x^n$ captures the suppressed spectrum distribution. According to the linear response theory, each frequency mode $x$ contributes to the heating rate separately, following Supplementary Eq.~\ref{eq.Floquetrate} as in Floquet systems. Therefore, to obtain the entire heating rates from all frequencies, one can simply perform a integral 
    \begin{equation}
        \Gamma = \int_0^{\infty}\!\mathrm{d}x\;g_x^2 A e^{-x\Omega/\epsilon}\propto \Omega^{-2n-1}(2n)!,
    \end{equation}
	 where we use the Gamma function, $F(n) = \int_0^{\infty} x^{n-1}e^{-x}\mathrm{d}x = (n-1)!$. Its inverse defines the heating time scale, $1/\Gamma \propto \Omega^{2n+1}$, which has been observed in our experiments. In fact, a generalized Floquet-Magnus expansion can be employed to investigate higher order processes. It goes beyond this simple linear response theory and still leads to the same prediction on the heating rates, see details in \cite{Mori2021}.

	 \subsection{Numerical results}
	 
	 We adopt two advanced tensor network algorithms, Projected Entangled Pair States (PEPS) and grouped Matrix Product States (GMPS), to simulate the dynamics generated by RMD protocols. Tensor networks are powerful methods to simulate quantum many-body physics and quantum circuits, providing highly efficient classical computational frameworks. Notably, they have demonstrated the capacity to surpass Google's quantum supremacy claims~\cite{pan2022simulation,pan2022solving}. Although tensor network algorithms are limited to simulating dynamics within constrained Hilbert spaces, they allow us to extract critical features from the numerical results. In this section, we initially present the Trotter-Suzuki decomposition, an effective method for approximately factoring time-evolution operators into products of local terms, thus making them suitable for tensor network simulations. Subsequently, we introduce these two methods, respectively, and compare their performance with our experimental observations. These numerical simulations will support our experimental observation, including $T^{2n+1}$ scaling in heating rates, oscillatory dynamics in entanglement entropy during the prethermal regime, as well as the rapid entanglement growth induced by heating.
	 
	 \subsubsection{Trotter-Suzuki Decomposition}
	 The time-dependent evolution operators are supported over the entire Hilbert space, applying directly to a quantum state is nearly infeasible. To address this challenge, we introduce the second-order Trotter-Suzuki decomposition \cite{paeckelTimeevolutionMethodsMatrixproduct2019a}. This method approximates the short-time evolution operator $e^{-i\hat{H}\delta t}$ as a product of operators with finite support, thereby facilitating practical computation and implementation. Specifically, we split the Hamiltonian as 
\begin{equation}
    \hat{H} = \hat{H}^1 + \hat{H}^2 + \cdots \hat{H}^m,
\end{equation}
where $\hat{H}^i$ is a sum of local operators $\hat{H}^i = \sum_{x,y}\hat{h}^i_{x,y}$. The second-order Trotter-Suzuki decomposition approximates the time-evolution operators as
\begin{equation}
\label{trotter-suzuki}
e^{-i \hat{H}\delta t} = \prod_{k=m}^1 e^{-i\hat{H}^k\delta t/2}\prod_{k=1}^m e^{-i\hat{H}^k \delta t/2} + \mathcal{O}(\delta t^3).
\end{equation}
The selection of $\hat{H}^k$ guarantees that each term $\hat{h}^k_{x,y}$ commutes pairwise. This commutativity implies that each operator on the right-hand side of Supplementary Eq.~\ref{trotter-suzuki} can be further decomposed into a product of local operators.
\begin{equation}
    e^{-i\sum_{x,y}\hat{h}^k_{x,y}\delta t/2} = \prod_{x,y}e^{-i\hat{h}^k_{x,y}\delta t/2}.
\end{equation}

We express our two-dimensional Hamiltonian as a sum of five distinct components.
\begin{align}
    \hat{H}^1_{\pm} &= \sum_{x,y} (1\pm \delta h)h_0[1-(-1)^y]a_{x,y}^\dagger a_{x,y} \label{eq:trotter_1}, \\
    \hat{H}^2 &= \sum_{x\text{ is odd}}\sum_{y}J  a_{x+1,y}^\dagger a_{x,y} + \mathrm{H.c.},\\
    \hat{H}^3 &= \sum_{x\text{ is even}}\sum_{y}J a_{x+1,y}^\dagger a_{x,y} + \mathrm{H.c.},\\
    \hat{H}^4 &= \sum_{x}\sum_{y\text{ is odd}}J  a_{x,y+1}^\dagger a_{x,y} + \mathrm{H.c.}, \\
    \hat{H}^5 &= \sum_{x}\sum_{y\text{ is even}}J a_{x,y+1}^\dagger a_{x,y} + \mathrm{H.c.}. \label{eq:trotter_5}
\end{align}
Here, $\hat{H}^1$ is the sum of single-site operators while $\hat{H}^2$ to $\hat{H}^5$ are the sum of two-site operators.

For extended time intervals $t$, we introduce timestamps $\delta t,\ 2\delta t,\ 3\delta t\dots$ and $T,\ 2T,\ 3T\dots$, and label them sequentially as $t_1, t_2,\dots$ in a chronological order, with $t_0 = 0$ and $t_N=t$. This construction guarantees that each time interval $t_{i+1}-t_i$ is less than $\delta t$ and ensures that the Hamiltonian remains time-independent.  

We use a raw tensor to estimate the Trotter errors, defined as the maximum discrepancy between the imbalance computed using the exact Hamiltonian and that obtained from an approximate Hamiltonian,
\begin{equation}\label{eq:trotter_error}
    \text{Trotter Error} = ||\mathcal{I}_{\text{raw}} - \mathcal{I}_{\text{exa}}||_\infty.
\end{equation}
Here $\mathcal{I}_{\text{raw}}$ is calculated using raw tensors in conjunction with the Trotter-Suzuki decomposition. We compute the Trotter errors exactly for the parameters $T=20$ ns and $n=2$ on $2\times 2$, $2\times 4$, $4 \times 2$ and $4\times 4$ lattices. The results are depicted in Supplementary Fig. ~\ref{fig:trotter_error}. In our subsequent simulations, we select $\delta t=3$ ns, such that the Trotter errors remain negligible.
\begin{figure}[h]
	\includegraphics[width=0.6\linewidth]{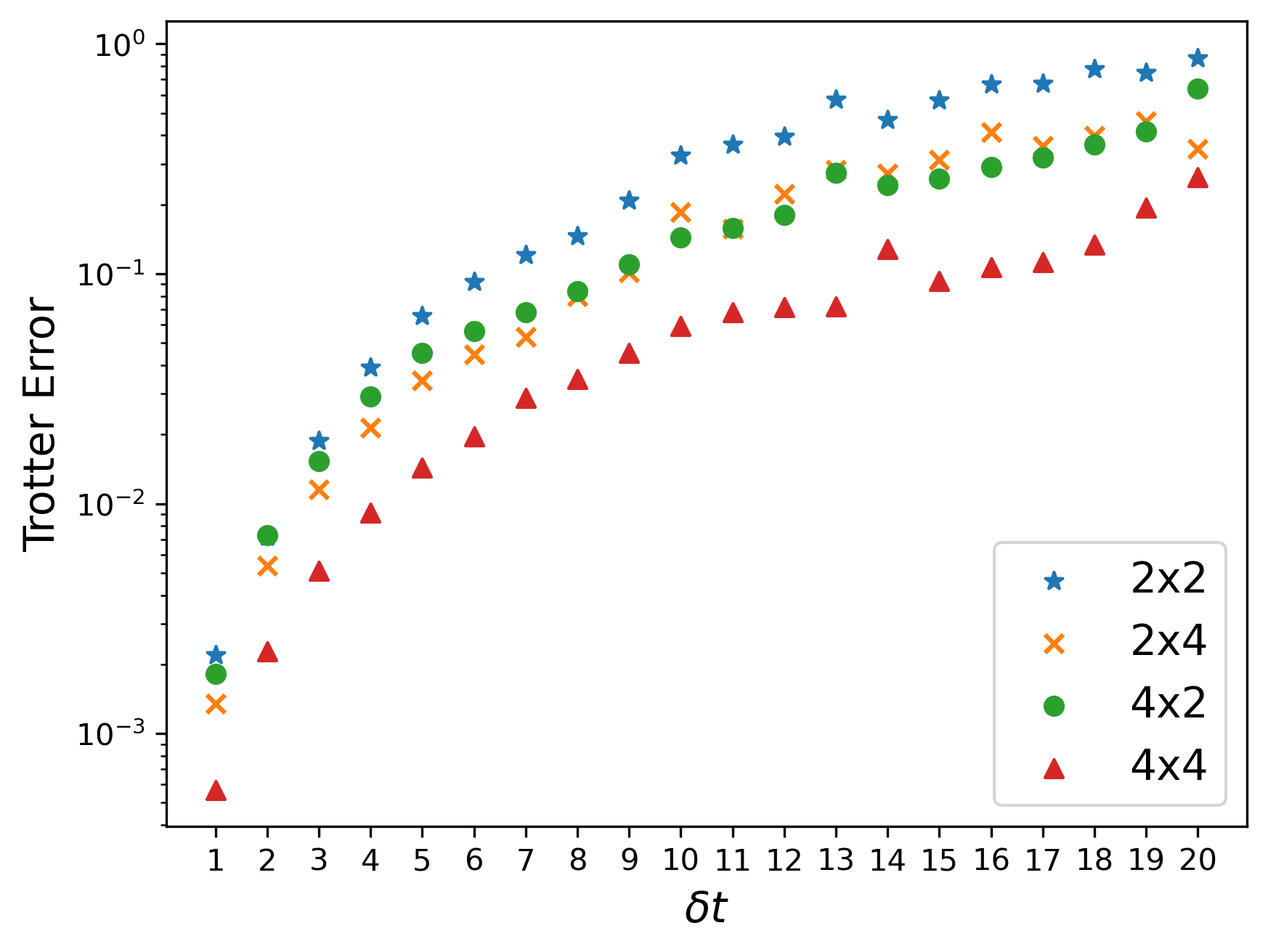}
	\caption{\textbf{Estimating Trotter errors ( Supplementary Eq.~\ref{eq:trotter_error}).} Evaluate the Trotter errors with different time steps on $2\times 2$, $2\times 4$, $4\times 2$ and $4\times 4$ lattices. The driving period is set to $T=20$ ns and $n=2$.}\label{fig:trotter_error}
\end{figure}
	 
	 \subsubsection{Project Entangled Pair States}
	 PEPS presents a way to parameterize many-body wave functions in a tensor network whose structure matches the two-dimensional underlying geometry of size $L_x \times L_y$. This approach is scalable in 2D and expressive, making it a potent tool for investigating the physics of strongly correlated systems \cite{verstraeterenormalizationalgorithmsquantummany2004, verstraetecriticalityarealaw2006} and for simulating quantum circuits \cite{liaosimulationibmskicked2023}.
	 
     We adopt the same decomposition strategy as Supplementary Eqs.~\ref{eq:trotter_1}-\ref{eq:trotter_5}. Choose the basis $\ket{0} \leftrightarrow \binom{1}{0}$ for the unoccupied state and $\ket{1} \leftrightarrow \binom{0}{1}$ for the occupied one. The factorized time-evolution operators in this basis are represented as
     \begin{equation}
         e^{-i\delta t h a^\dagger_i a_i} = \left[
         \begin{array}{cc}
            1  &  0\\
            0  &  e^{-i\delta t h}
         \end{array}\right].
     \end{equation}
     \begin{equation}
         e^{-i\delta t J(a^\dagger_ia_j+a^\dagger_ja_i)} = \left[
         \begin{array}{cccc}
            1 & 0 & 0 & 0\\
            0 & \cos(\delta t J) & -i\sin(\delta t J) & 0 \\
            0 & -i\sin(\delta t J) & \cos(\delta t J) & 0 \\
            0 & 0 & 0 & 1
         \end{array}
         \right].
     \end{equation}
     Single-site operators are $2\times 2$ matrices and thus can be directly applied to the single physical bonds. The two-site operators are $4\times 4$ full-rank matrices, which will increase the bond dimensions. We apply the two-site time evolution operators using simple update methods \cite{jiangaccurate2008,quthermal2019}, simply applying the operators on local tensors without accounting for extra environments, and then truncate to retain the largest bond dimension $\chi$. This approach ensures that the computational cost is managed while maintaining the essential features of the system's states.
     
     For generic PEPS, computing the expectation value of a single-site observable is computationally expensive due to the fact that fully contracting a PEPS is a $\#P$-hard problem. To address this, we employ the standard approximate method known as Time-Evolving Block Decimation (TEBD), which was originally designed to simulate the time evolution of matrix product states \cite{vidalEfficientSimulationOneDimensional2004, stoudenmire2010minimally}, and can also be utilized to contract two-dimensional PEPS. The computational complexity of this procedure scales as $\mathcal{O}(\chi^{12})$, making it the most resource-intensive step. The dynamics of imbalance from PEPS of $\chi=4$ are shown in Supplementary Fig.~\ref{fig:peps}. Each curve is obtained by averaging over ten different random realizations. Although PEPS captures the signal of prethermalization over a very short time scale, the systems experience accelerated thermalization during extended dynamic processes, manifesting as rapid decays of the imbalance. Furthermore, the observed scaling law of the thermalization lifetime is significantly shorter than theoretical predictions. Accurately modeling longer dynamics would incur a substantially higher computational cost compared to experiments.
     \begin{figure}
         \centering
         \includegraphics[width=1.0\linewidth]{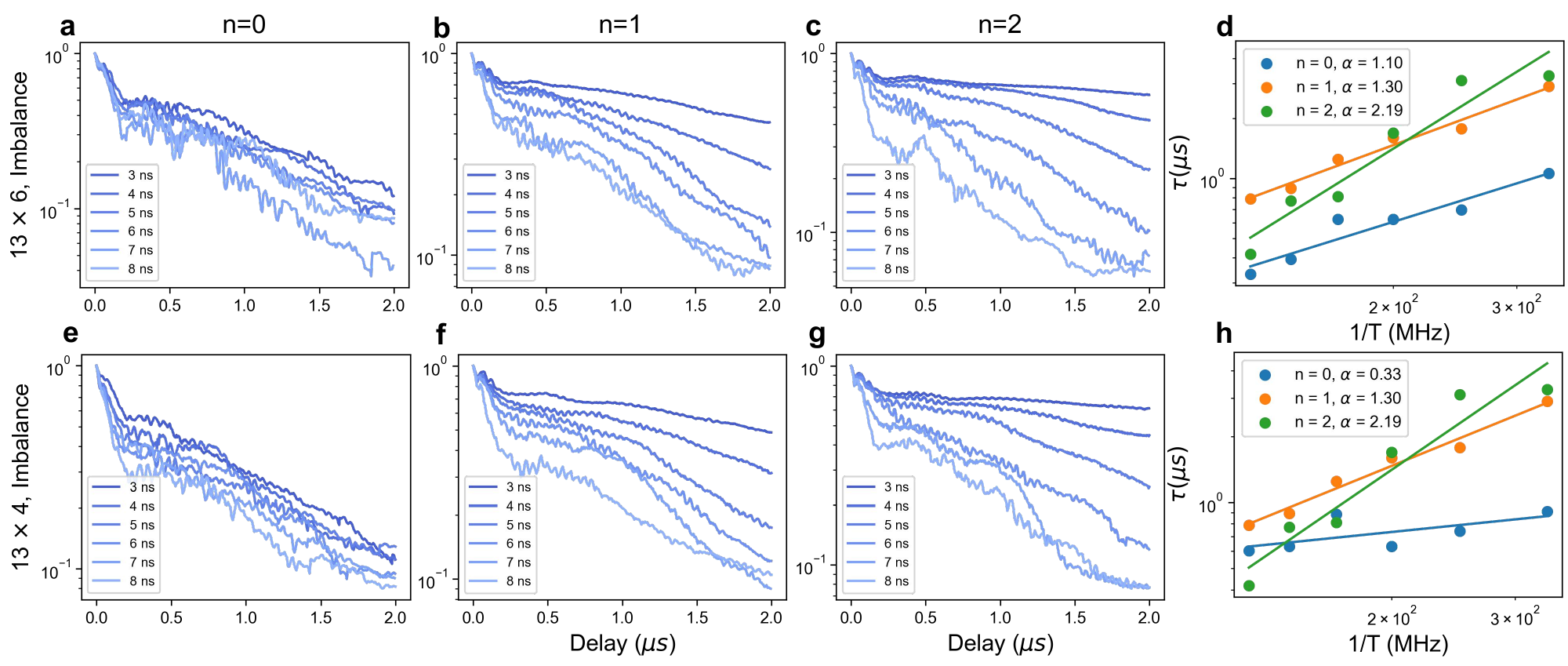}
         \caption{\textbf{The dynamics of imbalance in the system of size $13\times 6$ and $13\times 4$ obtained by PEPS.} The maximum bond dimensions are $\chi=4$. \textbf{a}-\textbf{c}, Size $13\times 6$, $n=0,1,2$ respectively, imbalance dynamics for $T$ ranging from 3 ns to 8 ns. \textbf{d}, We use a linear fitting to determine the power-law scaling, $\tau \propto T^{-\alpha}$,  in the log-log scale. \textbf{e}-\textbf{h}, Size $13\times 4$, same as \textbf{a}-\textbf{d}. All results are computed on Intel Xeon Platinum 8268 CPUs.}
         \label{fig:peps}
     \end{figure}
	 
	 \subsubsection{Grouped Matrix Product States}
	 Given the inherent limitations of PEPS methods, particularly the lack of a canonical form, we employ MPS. MPS are renowned for their efficiency and high accuracy in representing one-dimensional gapped quantum phases \cite{vidalEfficientSimulationOneDimensional2004}. With the availability of a canonical form, the expectation values of local operators can be computed by contracting only the local tensors, rather than the entire tensor networks.

     There are two possible MPS representations for a two-dimensional lattice. The first approach involves winding the MPS across the two-dimensional lattice, effectively forcing the system into a one-dimensional configuration. This transformation makes connections in the 2D lattice non-local, depending on the chosen 1D ordering. To apply non-local hopping operators in this configuration, additional SWAP gates are required to bring two sites into adjacency. We use the second approach, known as the grouped MPS method. In this approach, we group $L_y$ lattice sites in a single column, treating them as a single site with a physical dimension of $2^{L_y}$. This allows us to represent the system as a MPS of length $L_x$. The advantage of this method is that it preserves locality. However, the trade-off is that it is not scalable in the $y$-direction.
     
     We use the Trotter-Suzuki decomposition as follows
\begin{align}
    \hat{H}^1_\pm &= \sum_{x=1}^{L_x}\sum_{y=1}^{L_y}h_{\pm}a^\dagger_{x,y}a_{x,y} + J \left(a^\dagger_{x,y+1}a_{x,y} + \mathrm{H.c.} \right), \\
    \hat{H}^2 &= \sum_{x=1}^{L_y}\sum_{y=1}^{L_y}J \left(a^\dagger_{x+1,y}a_{x,y} +\mathrm{H.c.} \right).
\end{align}
The single-site operators are $2^{L_y}\times 2^{L_y}$ matrices and can be directly applied to each tensor without affecting the canonical position. For two-site hopping terms, we take $L_y=4$ and  $\hat{U} = \exp(-iJ\delta t(a^\dagger_{j=i+1,4}a_{i,4}+\mathrm{H.c.}))$ as an example and summarize update rules below. We also illustrate them in Supplementary Fig.\ref{fig:gmps}.

\begin{enumerate}
    \item Perform QR decompositions on each tensor to isolate the active physical bonds ($i_4$ and $j_4$ here) into $R$ tensors. This step is essential for ensuring that the calculations are in the most computationally efficient manner.
    \item Apply the two-site time evolution operator and contract into a single tensor.
    \item Execute a singular value decomposition (SVD) and eliminate the singular values that are smaller than a specified threshold. Subsequently, truncate the results by retaining the largest $\chi$ singular values.
    \item Absorb the singular values and vectors into the updated tensors.
\end{enumerate}

Owing to the canonical form in MPS, it is possible to monitor fidelity throughout the update process. During each truncation step, the singular values $s_1 \geq s_2 \geq \cdots \geq s_D$ are arranged in descending order, and only the top $\chi(<D)$ singular values are kept. Consequently, the fidelity is given by the ratio:
\begin{equation}
    \frac{\sum_{k=1}^\chi s_k^2}{\sum_{k=1}^D s_k^2},
\end{equation}
This expression provides a measure of how much of the original state's information is preserved after truncation. 

\begin{figure}
    \centering
    \includegraphics[width=0.9\linewidth]{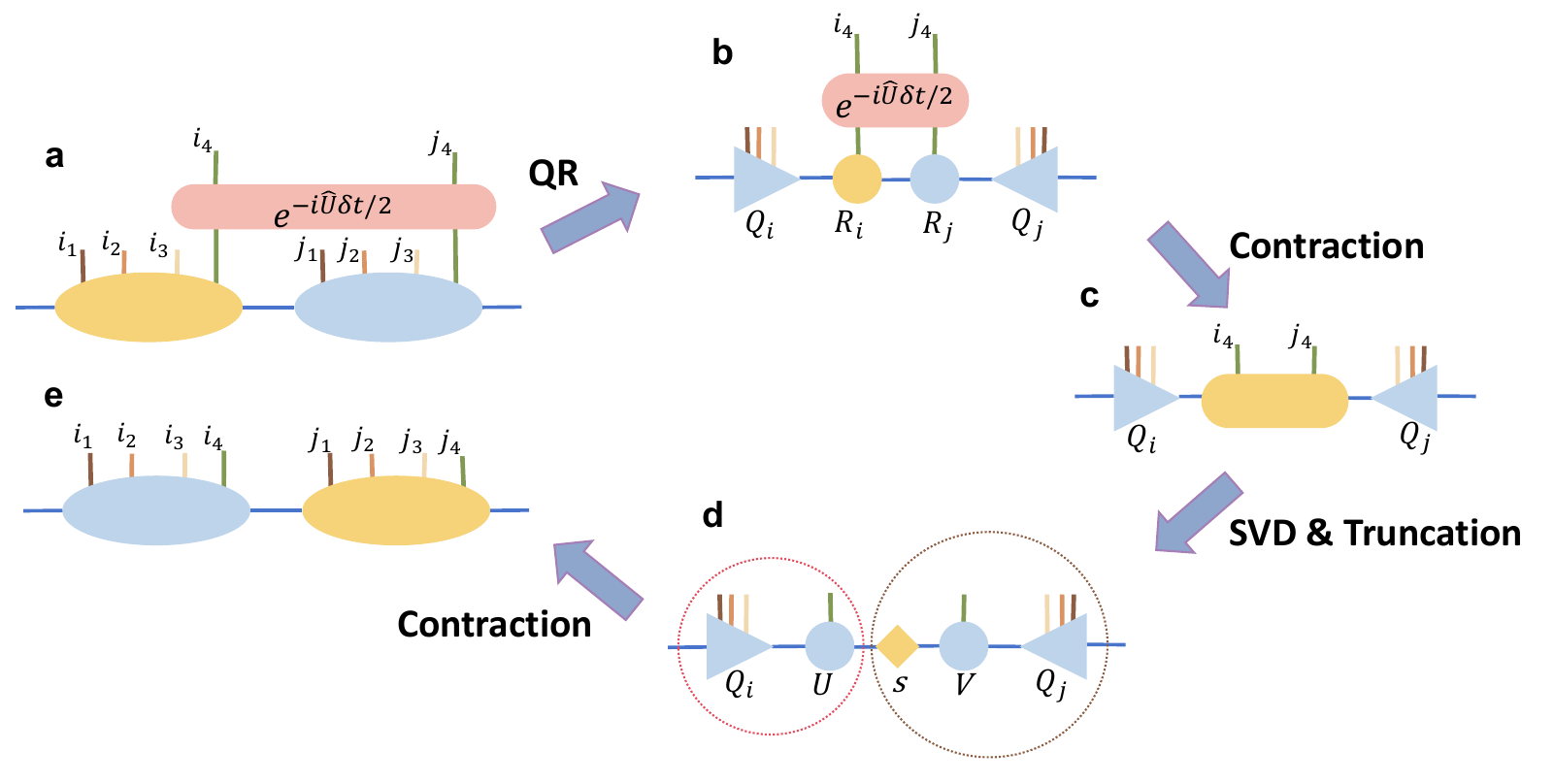}
    \caption{\textbf{A single step of time evolution under two-site operators.} $i_1\sim i_4$ and $j_1\sim j_4$ are labels for the local physical bonds along the $y$-direction. The canonical tensors are labeled in yellow. \textbf{a}-\textbf{b} QR decomposition on each tensor to extract the physical bonds $i_4$ and $j_4$ into $R_{i(j)}$ tensors. \textbf{b}-\textbf{c} Contract $R_{i(j)}$ tensors together with the time-evolution operator. \textbf{c}-\textbf{d} Apply singular value decomposition and keep the largest $\chi$ singular values. \textbf{d}-\textbf{e} Absorb singular values and vectors into $Q_{i(j)}$ to obtain the updated tensors.}
    \label{fig:gmps}
\end{figure}

The numerical results of the dynamics of the $13\times 6$ system are shown in Supplementary Fig.~\ref{fig:gmps_96}. All computations are performed on Intel Xeon Phi 7285 CPUs and it takes about ten hours to complete one trajectory simulation. We compute the imbalance (Supplementary Fig.~\ref{fig:gmps_96}\textbf{a}-\textbf{c}), the entropy of the upper leftmost site (Supplementary Fig.~\ref{fig:gmps_96}\textbf{d}-\textbf{f}), and the fidelity (Supplementary Fig.~\ref{fig:gmps_96}\textbf{g}-\textbf{i}). Each curve is obtained by averaging over ten trajectories. The results show that the imbalance exhibits exponential decay at short times ($t\lesssim 0.5$ ns, shown by the red dashed lines) and subsequently slows down. This time scale is also consistent with the deceleration as observed in the entropy growth. The short-time imbalance data are used to estimate the decay rate, as shown in Supplementary Fig.~\ref{fig:gmps_96}\textbf{j}, which exhibits the expected algebraic scaling $T^{2n+1}$ approximately. 

The decay of the estimated fidelity is also consistent with the heating process. During the prethermal regime, the system remains close to the density wave ordered state and the growth of the entanglement is slow. Therefore, MPS exhibits a high accuracy in numerical simulations. However, heating destabilizes the density wave order and at the same time, fidelity quickly drops. In particular, for $T=7$ ns and $n=2$ (corresponding to the orange experimental data in Fig.3), as shown in Supplementary Fig.~\ref{fig:gmps_96}\textbf{i}, a notable decay of the estimated fidelity is visible already at early times and it drops below 0.05 when $t{\approx} 1.5$~$\mu$s. This highlights that the non-equilibrium dynamics of a driven quantum system quickly become numerically intractable, especially for 2D systems like ours. 

\begin{figure}
    \centering
    \includegraphics[width=1.0\linewidth]{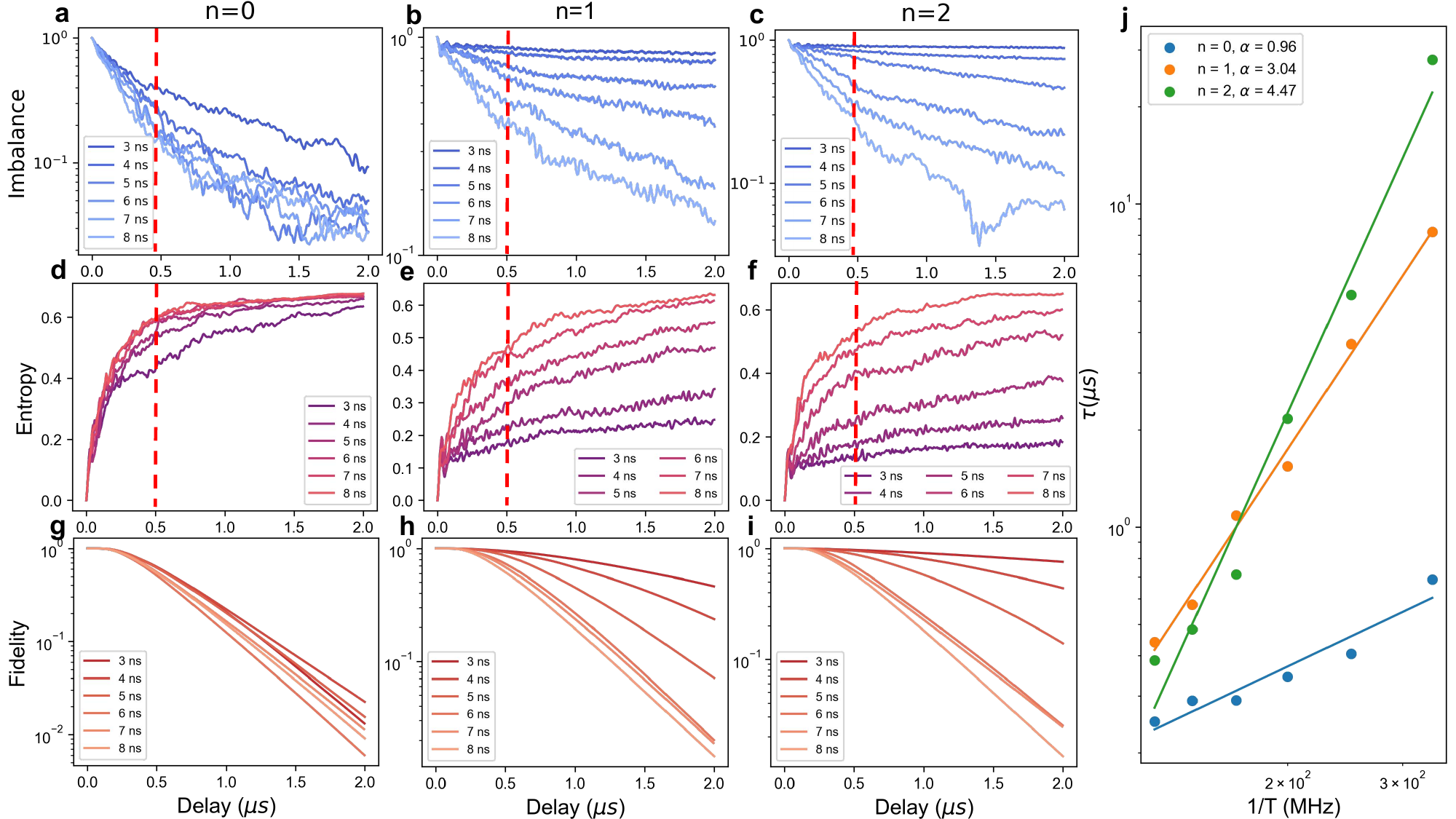}
    \caption{\textbf{Grouped MPS results for a $13\times 6$ system.} The maximum bond dimension is set to $\chi=96$, and we average ten random trajectories for each parameter being studied. \textbf{a}-\textbf{c}, Imbalance dynamics for various $T$, ranging from 3 ns to 8 ns, and $n=0,1,2$ respectively. \textbf{d}-\textbf{f}, Single-site entropy for various $T$ ranging from 3 ns to 8 ns, and $n=0,1,2$ respectively. The red dashed line approximately marks the time where noticeable errors appear. \textbf{g}-\textbf{i}, The corresponding fidelity of the MPS simulation. \textbf{j}, We employ the short-time imbalance dynamics ($t<0.5$ ns) to determine the lifetime $\tau$. It exhibits a power-law dependence $\tau \propto (1/T)^{\alpha}$. The slopes are $\alpha_{n=0}=0.96$, $\alpha_{n=1}=3.04$, $\alpha_{n=2}=4.47$, close to the theoretical prediction, 2n+1.}
    \label{fig:gmps_96}
\end{figure}

Utilizing the canonical form we can conveniently calculate the von Neumann entropy of a subsystem 
\begin{equation}
    S = -\text{Tr}\left(\rho \log \rho\right),
\end{equation}
with the reduced density matrix $\rho$. For instance, when calculating the von Neumann entropy of the subsystem $\mathcal{B}$ as depicted in Fig. 4\textbf{a} in the main text, all of the four qubits in $\mathcal{B}$ are in one column and grouped into the fifth MPS tensor from left to right. We begin by moving the canonical site to the fifth tensor. This arrangement ensures that when tracing out the tensors 1\textasciitilde 4 and 6\textasciitilde 13 tensors we obtain two identity matrices separately. Then we reshape the fifth tensor to a tensor of dimensions $(\chi, 2, \textbf{2}, \textbf{2}, \textbf{2}, \textbf{2}, 2,\chi)$ and modify it to a $\mathbf{16}\times 4\chi^2$ matrix $C$. Here, the bold numbers signify the physical dimensions of the subsystem $\mathcal{B}$. The reduced density matrix is $\rho_\mathcal{B} = CC^\dagger$ and von Neumann entropy can be calculated straightforwardly. For the subsystem $\mathcal{A}$, computation of the reduced density matrix is more complicated. We still benefit from the canonical form by tracing out the 1\textasciitilde 4 and 9\textasciitilde 13 tensors, but need to contract the 5\textasciitilde 8 tensors explicitly to get $\rho_\mathcal{A}$. For $\chi=96$, the evaluation of $S_\mathcal{A}$ takes approximately 6.9 seconds, whereas the evaluation of $S_\mathcal{B}$ only takes about 0.54 seconds on a personal computer.

For simplicity, we first analyse the dynamics of the entanglement entropy of the top leftmost single site, as shown in Supplementary Fig. ~\ref{fig:gmps_96}\textbf{d}-\textbf{f}. Entanglement entropy grows at different rates for different multipolar order and the driving period, $T$. Crucially, for multipolar order $n\geq1$, the increase of entanglement entropy can be substantially suppressed by using a fast drive, highlighting the importance of the temporal correlation in stabilizing the system. However, heating is still inevitable and leads to notable entanglement generation at later times, significantly increasing the computational resources in simulating the time evolution.

\begin{figure}[h]
    \centering
    \includegraphics{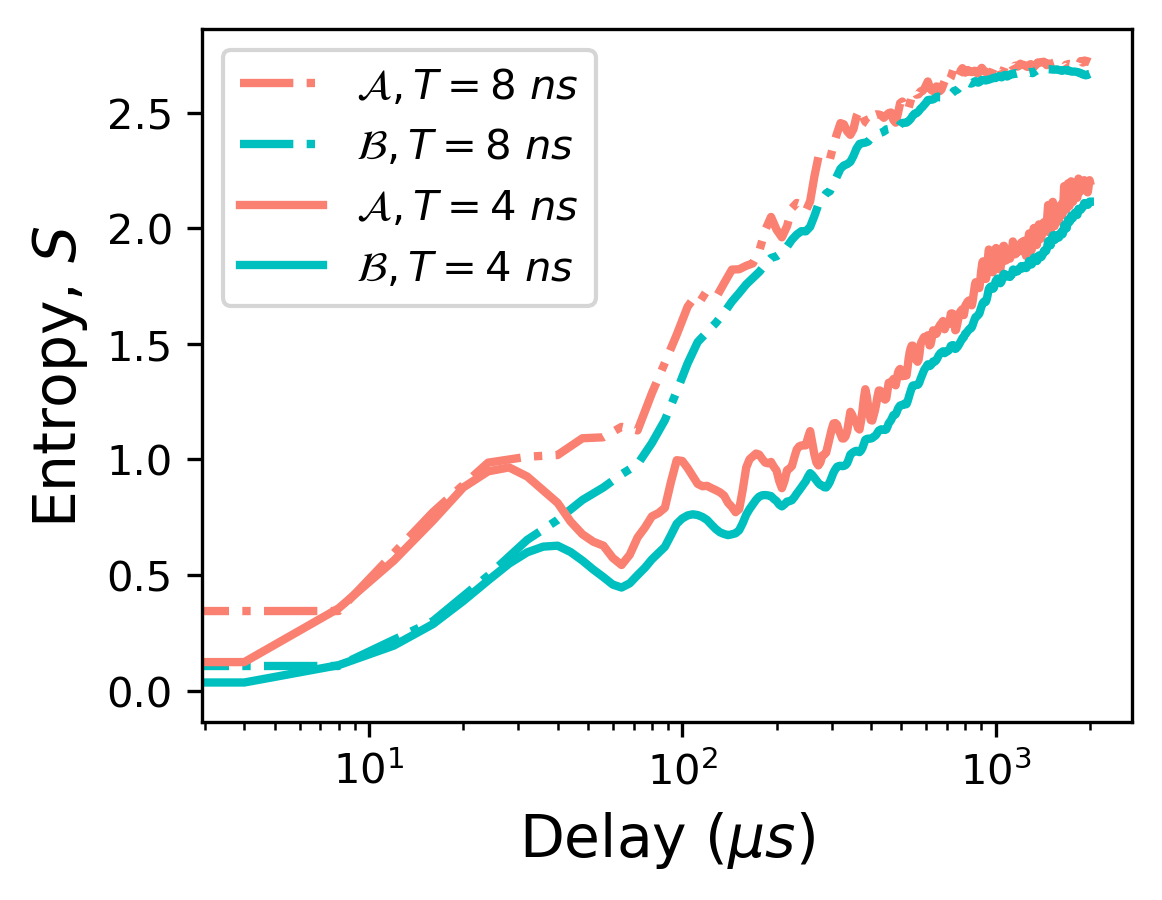}
    \caption{\textbf{Entanglement dynamics.} The dynamics of entanglement entropy for different subsystems $\mathcal{A}$ and $\mathcal{B}$. $\mathcal{A}$ and $\mathcal{B}$ both exhibits oscillatory behaviors when $T = 4$ ns. In the experiment, however, we only observe the pronounced oscillatory dynamics in subsystem $\mathcal{A}$. For subsystem $\mathcal{B}$, due to the presence of noise and decoherence, as shown in Fig.4\textbf{a}, its entanglement entropy quickly saturates at a prethermal plateau in real experiments. We use 1-RMD drives.}
    \label{fig:xy_entropy}
\end{figure}
We further analyze the entanglement dynamics of subsystems $\mathcal{A}$ and $\mathcal{B}$, as illustrated in Supplementary Fig.~\ref{fig:xy_entropy}. For a rapid drive with $T = 4$ ns, our numerical simulation successfully reproduces the oscillation behavior of entanglement within the prethermal regime. It is worth noting that, the subsystem $\mathcal{A}$ exhibits a more pronounced oscillation amplitude than the subsystem $\mathcal{B}$, indicating a non-uniform entanglement distribution in this prethermal plateau. This behavior has also been experimentally detected, as shown in Fig.4\textbf{a} in the main text. However, due to the presence of noise and decoherence, the oscillatory dynamics in subsystem $\mathcal{B}$ becomes invisible experimentally. Also, for a slower drive (dashed dotted line in Supplementary Fig.~\ref{fig:xy_entropy}), the system heats up swiftly without exhibiting the prethermal plateau.

\begin{figure}[h]
    \centering
    \includegraphics[width=\linewidth]{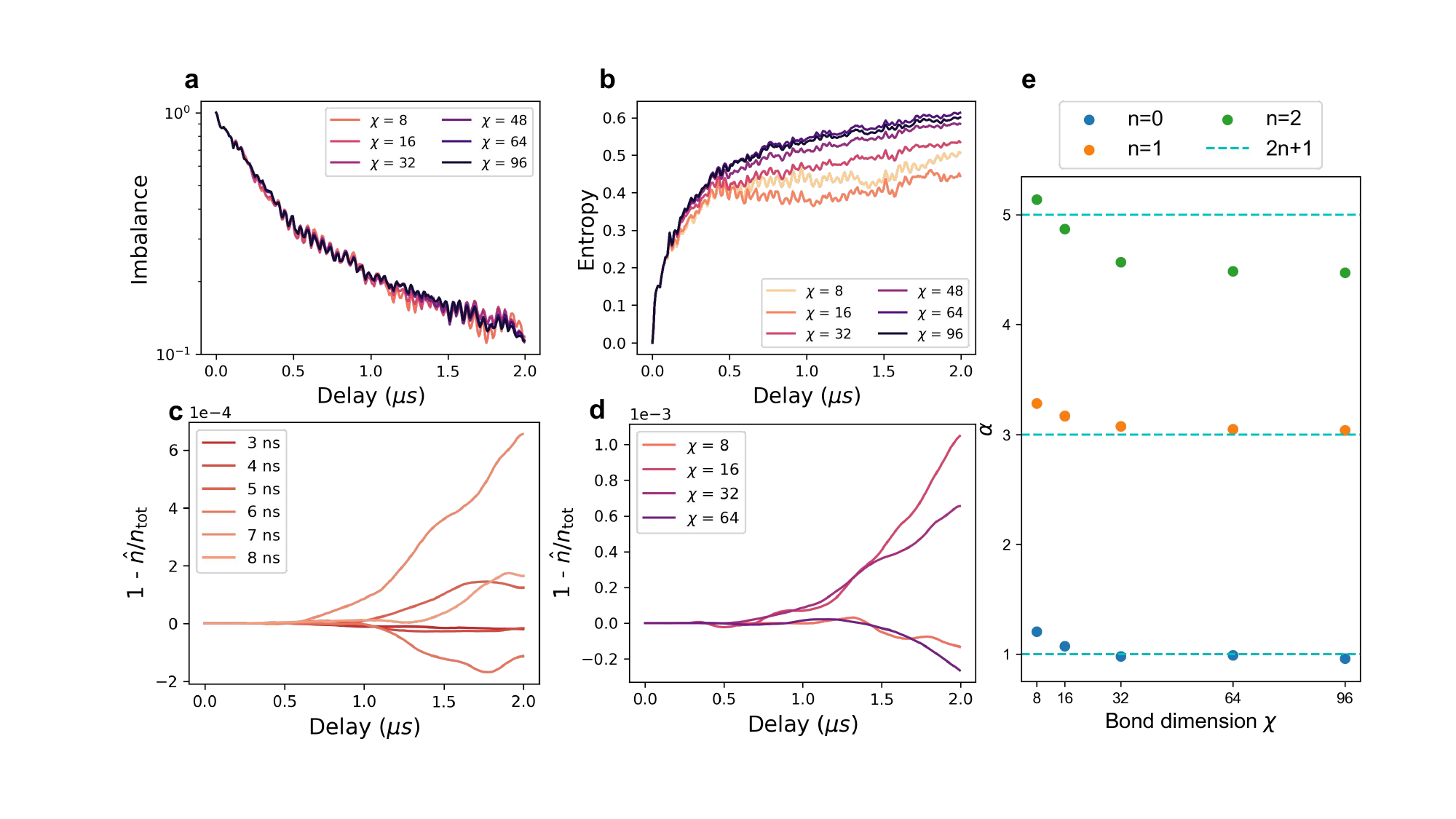}
    \caption{\textbf{Grouped MPS with varying maximum bond dimensions.} \textbf{a}-\textbf{b}, Depicts the imbalance and single-site entropy for a $13\times 6$ lattice, the time period is $T=7$ ns and with 2-RMD drives. \textbf{c}, Particle numbers with 2-RMD drives and different periods. The relative errors are at scale $10^{-4}$ and the particle numbers are well preserved. \textbf{d}, Particle numbers at different maximum bond dimension, the time period is $T=7$ ns and with 2-RMD drives. \textbf{e}, Displays the exponents $\alpha$ obtained through fitting with imbalance across various bond dimensions $\chi$; the dots represent the fitted values, while the dashed lines correspond to $2n+1$.}
    \label{fig:gmps_convergence}
\end{figure}
Moreover, we investigate the performance of grouped MPS with different maximum bond dimensions. In Supplementary  Fig.~\ref{fig:gmps_convergence}\textbf{a}-\textbf{b}, the congruence of these imbalance and entropy is sustained only for a short time, beyond which discrepancies appear at later times. Varying values of $\chi$ also affects the largest reachable entanglement entropy. Even with $\chi=96$ the entropy is still far from the Page value at long times, suggesting the computational challenge in simulating the heating process. 
We calculate the scaling exponents by extracting the prethermal lifetime for different $T$ based on the imbalance decay and present the results in Supplementary Fig.~\ref{fig:gmps_convergence}\textbf{e}. For $n=0$ and $1$ the exponents converge to the theoretically prediction $\alpha=2n+1$(dashed lines) when $\chi \geq 32$. However, for $n=2$, the exponent converges to around 4.5 which is still smaller than $5$. It happens because for large $n$, typically one needs an extremely long time scale to extract the scaling exponent, as shown in \cite{ZhaoRandom2021}, which is far beyond the reach of MPS simulations. In fact, similar behavior also appears in our experiments. As shown in Fig.3 \textbf{a}, due to the presence of noise and decoherence, for large system sizes the scaling exponent for $n=2$ is around 3.6.

We also check the conservation of total particle number
\begin{equation}
    \hat{n} = \sum_{x,y} a_{x,y}^\dagger a_{x,y},
\end{equation}
during the MPS simulation. It
remains almost conserved at the value $n_\text{tot} = L_xL_y/2$ throughout the driven dynamics.
This quantity serves as a critical criterion for the performance of our algorithms. Our analysis reveals that the relative errors in particle numbers are around $10^{-4}$ (Supplementary Fig.~\ref{fig:gmps_convergence}\textbf{c}), indicating that convergence is robustly achieved. Furthermore, this conservation is consistently maintained across different values of $\chi$ (Supplementary Fig.~\ref{fig:gmps_convergence}\textbf{d}).

In summary, we perform extensive numerical simulations via tensor network techniques. GMPS exhibits a better performance than PEPS for our systems. GMPS is capable of capturing the early time dynamics, based on these data we confirm the $T^{2n+1}$ in the suppressed heating rates. However, the appearance of heating accelerates the growth of the entanglement entropy and GMPS method requires significant amount of computational resources in emulating our 78-qubit experiments over 1,000 driving cycles. Therefore, our experiment potentially demonstrates a quantum advantage in studying the non-equilibrium many-body dynamics in driven systems.

    \clearpage
    \newpage

	\section{Experimental setup}
    Our experiments are performed on a superconducting quantum processor, named \textit{Chuang-tzu 2.0}, comprises 78 qubits arranged in a square lattice configuration with 6 rows and 13 columns. Each pair of nearest-neighbor qubits is interconnected by an adjustable coupler, resulting in a total of 137 couplers within the processor. In Supplementary Fig.~\ref{WiringInfo}, we show the room temperature and cryogenic wiring layout. The cryogenic environment, maintained at approximately 12 mK, is achieved using a BlueFors dilution refrigerator (model XLD1000). A carefully designed arrangement of attenuators and filters has been implemented to effectively suppress thermal noise, ensuring precise control of qubits and couplers.  
    
    To meet the high-frequency requirements of RMD, we utilize a dual-channel digital-to-analog converter (DAC) operating at 2 Gs/s with 16-bit vertical resolution and dual-channel analog-to-digital converter (ADC) at 1 Gs/s with 14-bit vertical resolution, controlled by a field-programmable gate array (FPGA) chip. This system, functioning as the Z control for both qubits and couplers, is capable of generating nanosecond-scale pulses, ensuring the precision and speed necessary for advanced quantum operations. Moreover, when combined with a microwave source acting as the local oscillator (LO), the two DAC channels can modulate the in-phase and quadrature components of an IQ mixer, enabling frequency up-conversion to generate signals in the GHz range. These signals are critical for qubit XY control and dispersive readout operations. We also perform zero calibration of each IQ mixer to reduce intrinsic and mirror leakages. The custom DAC/ADC devices are synchronized using a trigger signal and a 250 MHz reference clock. Communication between the host computer and the devices is achieved through a high-speed local area Ethernet network, leveraging fiber-optic connections with a bandwidth of 10,000 Mbps.
	
	\begin{figure}
		\centering
		\includegraphics[totalheight=5in]{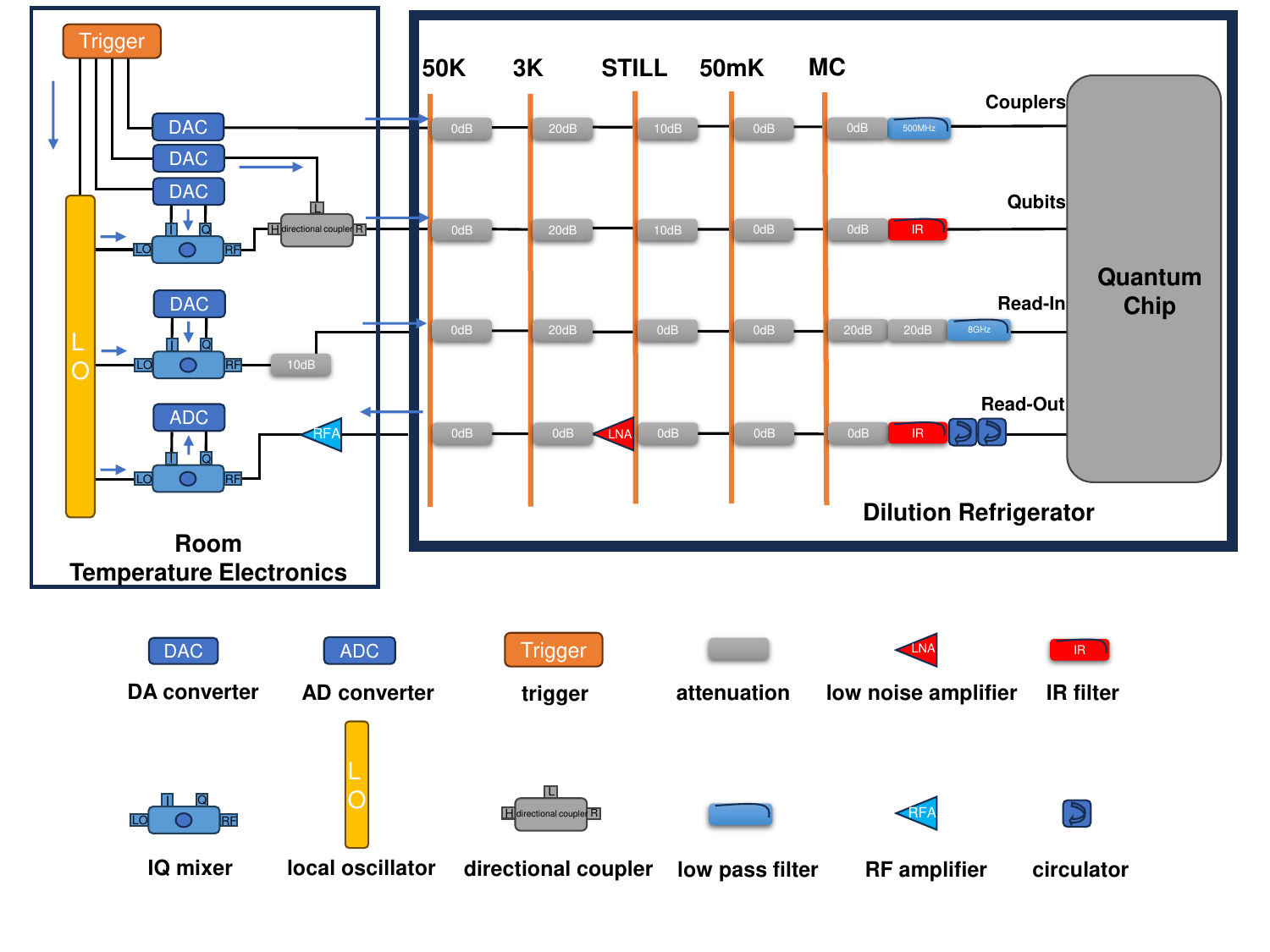}
		\caption{\textbf{Room temperature and cryogenic wiring.}} \label{WiringInfo}
	\end{figure}
    
    \section{Device fabrication and performance}
	\subsection{Device fabrication}
	The optical micrograph of the processor is displayed in Supplementary Fig.~\ref{optical_micrograph}.
	It is noteworthy that the qubit connectivity in our processor, while maintaining a square lattice structure, differs slightly from processors such as Sycamore~\cite{Arute2019}, which feature a square lattice with zig-zag edges.
    We employ a tunable coupler with a capacitively connecting pad (TCCP) architecture~\cite{Liang2023}, which facilitates a 1200-$\mu$m spacing between adjacent qubits, thereby ensuring sufficient wiring space. The qubits and couplers are fabricated on the qubit layer chip, whereas the control lines, readout lines, and readout cavities are integrated into the wiring layer chip. These two chips are interconnected using flip-chip bonding technology. The readout cavities for the 6 qubits in each column are multiplexed onto a single readout line and are capacitively coupled to the qubits through interfacial capacitance. The control lines are similarly coupled to the qubits and couplers, enabling excitation and biasing via interfacial capacitance and mutual inductance.
    
    \begin{figure}
		\centering
		\includegraphics[width=0.97\linewidth]{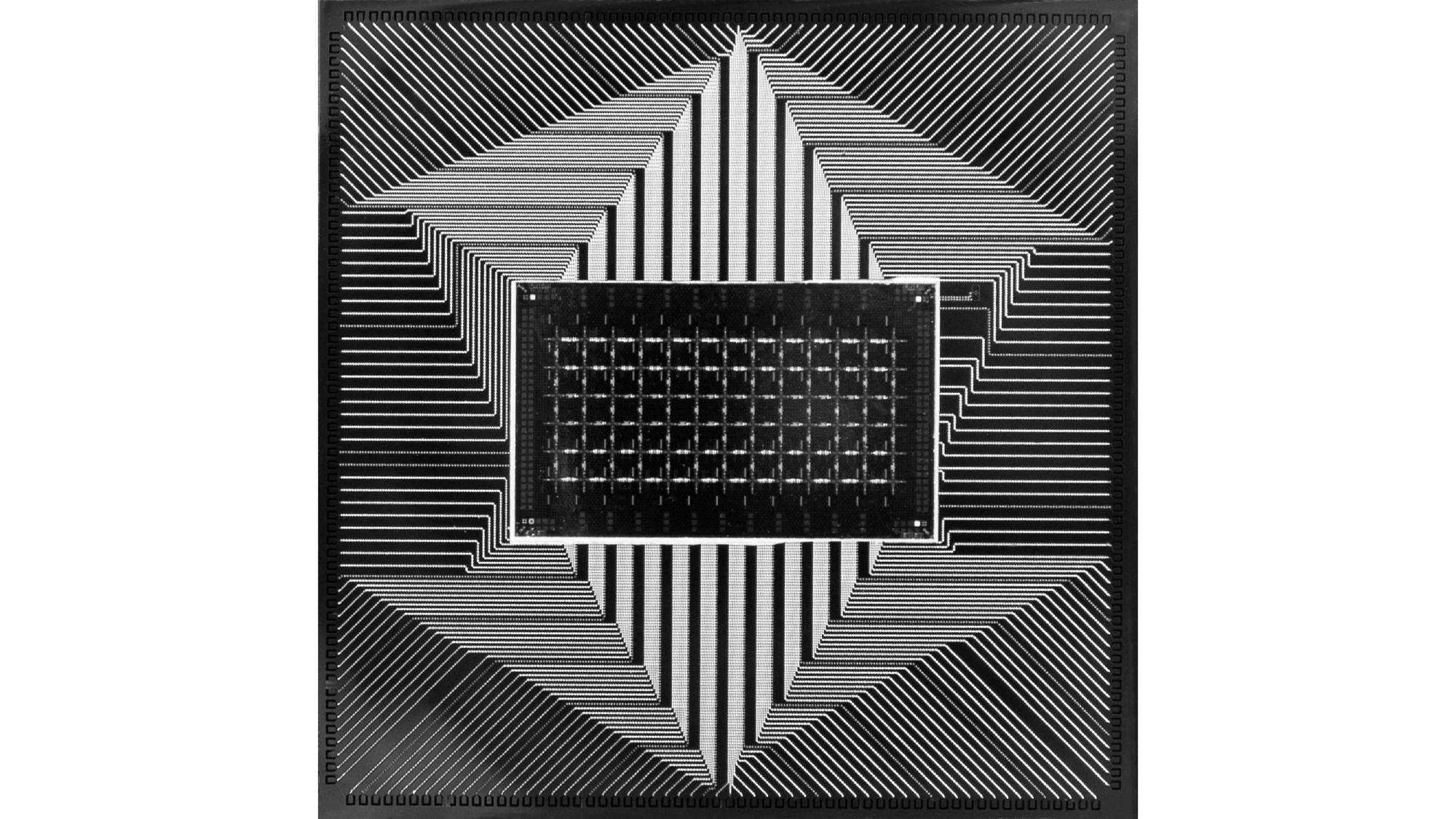}
		\caption{\textbf{Optical micrograph of the processor \textit{Chuang-tzu 2.0}.}} \label{optical_micrograph}
	\end{figure}
	
    % The processor  consists of 78 qubits in 6 rows and 13 columns, which are arrayed as a square lattice. It may be noticed that the connectivity of qubits as a square lattice of our processor is slightly different from the processors like Sycamore which have a square lattice configuration but with zig-zag edges. Each pair of nearest neighbor qubits has an adjustable coupler between them, with a total of 137 couplers in this processor. 
    % We utilize tunable coupler with capacitively connecting pad (TCCP) architecture~\cite{Liang2023}, allowing a spacing of 1200-$\mu$m between two qubits, thus providing ample wiring space. The qubits and couplers are designed on the qubit layer chip, while the control lines, readout lines, and readout cavities are designed on the wiring layer chip. The two chips are connected using flip-chip bonding. The readout cavities for the 6 qubits in each column share a single readout line, and are coupled to the qubit through interfacial capacitance. The control lines are coupled to the qubits and couplers for excitation and biasing through interfacial capacitance and mutual inductance.
        
    The superconducting qubit processor utilized in the experiment was fabricated using an aluminum film deposited on a sapphire substrate, combined with a flip-chip bonding process. The detailed fabrication procedure is outlined as follows:
    \begin{enumerate}
        \item Two two-side polished sapphire wafers are annealed by electron beam evaporation equipment. Then 100-nm aluminum is evaporated, followed by in-situ oxidation to generate a dense aluminum oxide. This dense aluminum oxide can both protect the aluminum film and make subsequent etching processes more uniform.
        \item Apply SPR-955 photoresist to the wafer surface. Use laser direct writing to expose patterns of the qubit layer and wiring layer on two wafers respectively. Develop with Tetramethylammonium hydroxide solution and etch with heated one. Finally, use the N-Methylpyrrolidone to remove the photoresist to obtain two the qubit wafer and wiring wafer.
        \item We need to prepare niobium on the wafers as a transition layer between aluminum and indium, as well as for the alignment mark of the electron beam exposure. Spin-coat LOR-5A and SPR-955 photoresists sequentially, then expose and develop to obtain patterns with undercut structures. Use magnetron sputtering to in-situ argon ion cleaning to remove the aluminum surface aluminum oxide layer and then deposit 100-nm niobium. Finally, remove the photoresist and lift-off the excess niobium film from the surface of the wafers.
        \item Prepare Josephson junctions on the qubit wafers. Spin-coat MAA-EL9 and PMMA-A5 bilayer resist sequentially and then deposit 10-nm aluminum for conductivity. Perform multiple exposures with different doses using electron beam lithography, and then remove the surface 10-nm aluminum, and develop using MIBK solution diluted with isopropanol. Employ in-situ argon ion cleaning with electron beam evaporation to remove the surface aluminum oxide layer, followed by double-angle evaporation with 65-nm and-100 nm aluminum respectively. Introduce in-situ oxidation between the two evaporation steps to form the insulating region in the Josephson junction's two-layer superconductors. By controlling the oxidation time, the Josephson junction's resistance and Josephson energy can be managed. Finally, remove the photoresist and lift-off the excess aluminum film, and its resistance is tested using a probe station to ensure it meets the design specifications.
        \item Both the qubit wafer and wiring layer wafer need to prepare air-bridges to achieve better ground sharing and reduce various parasitic modes. Start by spin-coating SPR-220 photoresist, exposing and developing the pillar patterns for the air-bridges. Then perform reflow to create arched air-bridges surfaces. Use electron beam evaporation to deposit 500-nm aluminum after in-situ argon ion cleaning, followed by in-situ oxidation. Spin-coat S-1813 photoresist, expose and develop the areas outside the air-bridges, and then etch using heated aluminum etchant type-D. Clean with pure oxygen using reactive ion etching and  then remove the resist to obtain the air-bridges.
        \item Prepare indium bumps for mechanically and superconductively connecting the two chips. Spin-coat AZ-4620 photoresist for the first exposure to create the undercut region. Wait for a period to allow the photoresist to rehydrate and degas, then spin-coat S-1813 photoresist for the second exposure in the indium bump area. After development, a structure with a deep undercut can be obtained. Use a indium thermal evaporation equipment to deposit 7-$\mu$m of indium after in-situ argon ion cleaning. Clean with a reactive ion etching using a mix of argon and oxygen, and remove the photoresist and lift-off the excess indium film to obtain the indium bumps.
        \item Finally, we will dice the wafers into chips and proceed with the packaging. Spin-coat AZ-4620 to protect the wafer, then use a dicing saw to cut the wafer into chips for the qubit layer (11mm$\times$18mm) and wiring layer (35mm$\times$35mm). After removing the photoresist, the two chips are bonded using flip-chip bonding equipment. Lastly, the chips are packaged into sample boxes using an ultrasonic wire bonder. After verifying that all connections from the sample box to the chip are functioning correctly using a multimeter, the chip can undergo low-temperature testing.
    \end{enumerate}
	
	\subsection{Device performance}
    In Supplementary Fig.~\ref{fig_devcice_performance}, we display the basic performance of qubits, including mean energy relaxation time $\bar{T}_1$, Ramsey dephasing time at idle frequency $T_2^*$, qubit readout frequency $f_r$, idle frequency $f_{10}$, and single-qubit gate error. The flux crosstalk between neighbouring qubits is below $10^{-3}$. The hopping strength between neighboring qubits can be precisely adjusted using couplers, with a tunable range spanning from -10 MHz to 3 MHz. This flexibility allows for straightforward achievement of the target value, $J/2\pi=2$~MHz, Notably, the average anharmonicity is $U/2\pi=-200$~MHz, with a ratio $|U/J|\approx 100\gg 1$, our processor can be regarded as a hard-core bosonic system~\cite{yan_strongly_2019}. All measured qubit probabilities are corrected to mitigate the readout errors.
    
    \begin{figure}[H]
		\centering
		\includegraphics[width=0.97\linewidth]{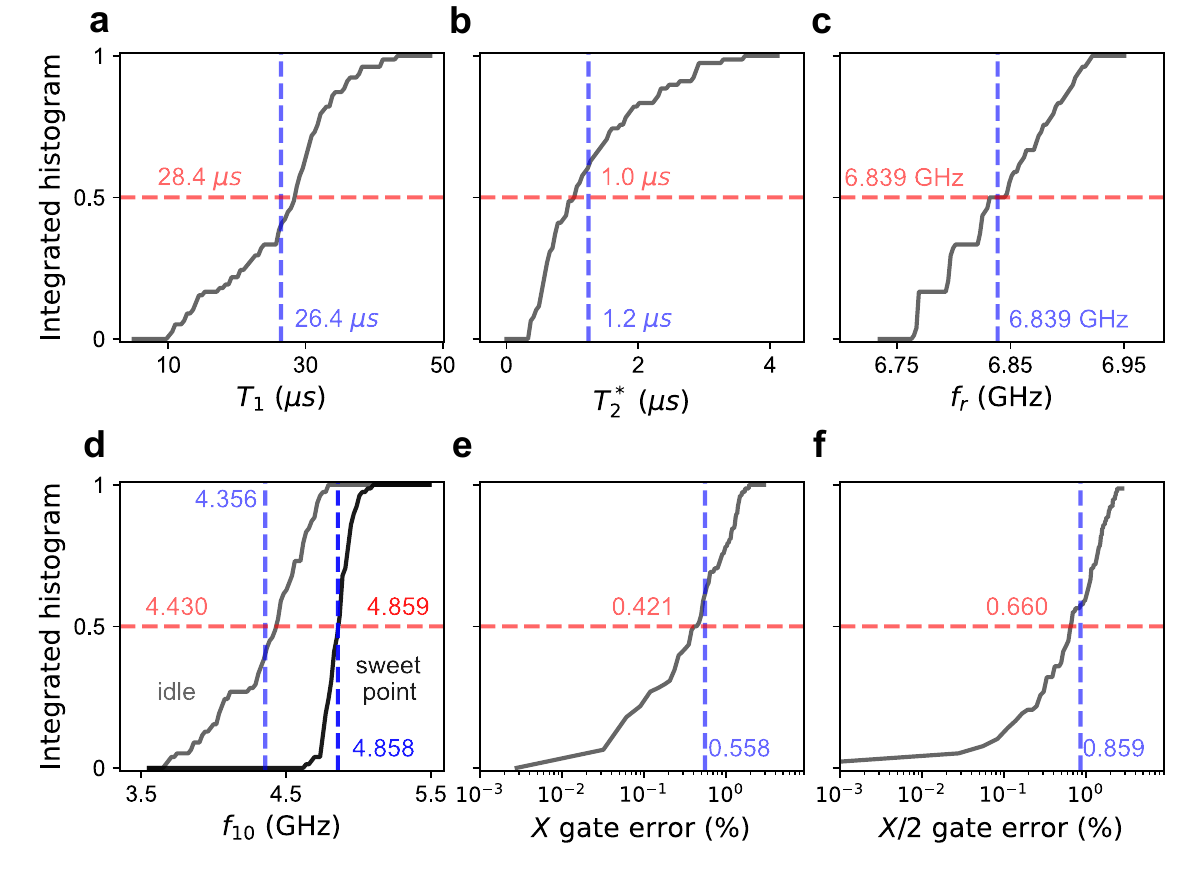}
		\caption{\textbf{Device performance of the processor \textit{Chuang-tzu 2.0}.} \textbf{a}, Distribution of the mean energy relaxation time $\bar{T}_1$. \textbf{b}, Distribution of the dephasing time $\bar{T}_2^*$ measured by Ramsey experiments at idle points. \textbf{c}, Distribution of the readout frequency $f_r$. \textbf{d}, Distribution of qubit frequency of sweet points (the darker curve) and idle points (the lighter curve). \textbf{e}, Distribution of the $X$ gate error. \textbf{f}, Distribution of the $X/2$ gate error. The single-qubit gates are characterized by quantum process tomography (QPT).  In these panels, the blue vertical dashed lines indicate the mean values, while the red horizontal dashed lines represent the median values.} \label{fig_devcice_performance}
	\end{figure}

    \clearpage
    \newpage
	
	\section{Device control and calibration}
	\subsection{Pulse calibration}
	When signals propagate from the DAC device to the qubits and couplers, the pulse shape undergoes distortion arising from the finite bandwidth limitations of various electrical components. The distortion of the Z pulses can induce an unintended frequency drift prior to the readout process. In our experiments, we implement precise correction protocol to effectively compensate for the distortion of Z pulses.
	
	Typically, the measurement circuit can be modeled as a linear time-invariant (LTI) system, where the output is fully determined by its impulse response $h(t)$:
	\begin{equation}\label{trans1}
	    y(t)=S[x(t)]=\int_{-\infty}^{+\infty}\!\mathrm{d}\tau\; x(\tau) h(t-\tau)=x(t)*h(t),
	\end{equation}
	where $f*g$ denotes the convolution between $f$ and $g$, $h(t)$ represents the system's output when subjected to a Dirac delta function $\delta (t)$ as the input
	\begin{equation}
	    h(t)=S[\delta(t)].
	\end{equation}
    Alternatively, the unit step response of the Z pulse during propagation can be measured, i.e., $s(t)=S[\theta(t)]$, with the unit step function $\theta(t)$, offering a more practical and convenient implementation, and $h(t)$ is given by
    \begin{equation}\label{supp_eqht}
        h(t)=\dot{s}(t).
    \end{equation}
    In frequency domain, we have Fourier transformations $Y(j\omega)=\mathcal{F}[y(t)]$, $H(j\omega)=\mathcal{F}[h(t)]$ and $X(j\omega)=\mathcal{F}[x(t)]$, such that Supplementary Eq.~\eqref{trans1} is transformed as
    \begin{equation}
        Y(j\omega)=H(j\omega)X(j\omega),
    \end{equation}
    with $H(j\omega)\approx 1$. Thus, the traget signals can be effectively reconstructed through deconvolution with the inverse of the frequency response. According to Supplementary Eq.~\eqref{supp_eqht}, the key procedure for distortion correction is to obtain the unit step response $s(t)$. The pulse sequences for correcting the Z pulse
    distortion in qubits and couplers~\cite{Li2025} is shown in Supplementary Fig.~\ref{pulseshape}\textbf{a} and \textbf{b}, respectively. 
    
    To correct the distortion in a qubit, a long pulse with duration about 3~$\mu$s and and a fixed amplitude is applied to the Z channel, generating a step signal. After a specific delay, a $\pi$-pulse is applied to the qubit, which can only excite the qubit to the $\ket{1}$ state when its frequency matches that of the $\pi$-pulse. The probability of the qubit being in the $\ket{1}$ state is then measured, as illustrated in Supplementary Fig.~\ref{pulseshape}\textbf{c}. By determining the peak population of the  $\ket{1}$-state for each delay, the Z offset data can be fitted using an exponential decay function. By employing the deconvolution method, we generate a corrected waveform that is subsequently transmitted to the arbitrary waveform generator (AWG). As a result, the qubit experiences a significantly flattened Z pulse, as demonstrated in Supplementary Fig.~\ref{pulseshape}\textbf{e}.
    
    In addition, the Z distortion in the coupler can be effectively corrected by exploiting its interaction with the neighboring qubit. This is achieved by first applying a step signal to the coupler, followed by a $\pi$-pulse to the neighboring qubit after a specific delay. The qubit and coupler are then biased into resonance to enable efficient energy exchange by a fine-tuned swapping duration and amplitude. Following this swapping process, the probability of the $\ket{1}$ state is measured. Similarly, the population of the $\ket{1}$ state is maximized only when the coupler frequency precisely matches the qubit frequency. As demonstrated in Supplementary Fig.~\ref{pulseshape}\textbf{d} and \textbf{f}, the peak probability of the 
  $\ket{1}$ state can be fitted to accurately characterize and correct the Z distortion in the couplers.
	
	\begin{figure}
		\centering
		\includegraphics[width=1.02\linewidth]{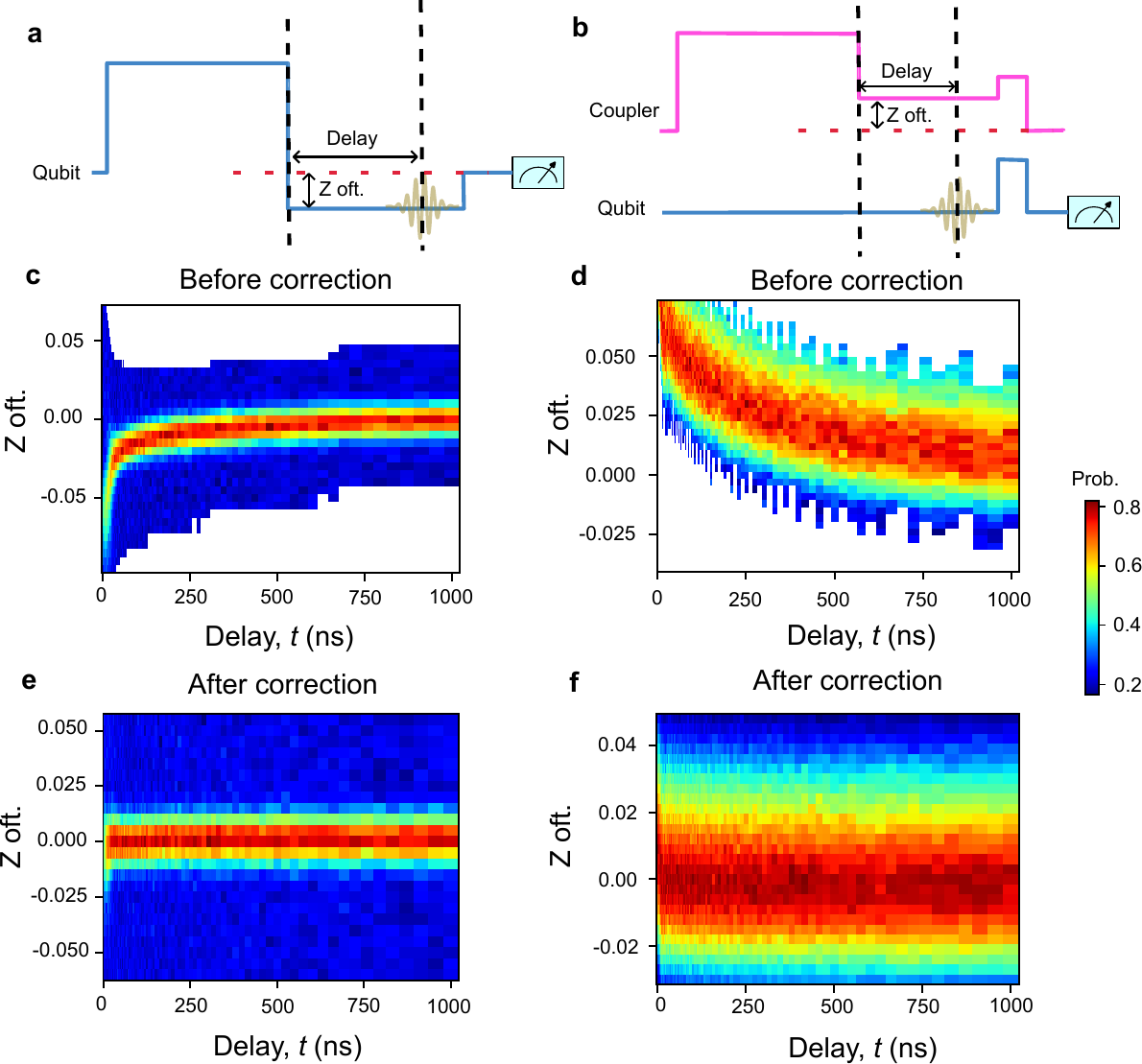}
		\caption{\textbf{Calibration of the Z pulse distortion for qubits and couplers.} \textbf{a}, Schematic of the experimental pulse sequence for correcting Z disortion of qubits. \textbf{b}, Schematic of the experimental pulse sequence for correcting Z disortion of couplers. \textbf{c} and \textbf{d}, The $\ket{1}$ state probability before correction for qubits and couplers, respectively.
		\textbf{e} and \textbf{f}, The $\ket{1}$ state probability after correction for qubits and couplers, respectively.}
		\label{pulseshape}
	\end{figure}
	
	\clearpage
	\newpage
	
	\subsection{The RMD pulse}
    To implement the Hamiltonian described in the main text, the qubit frequency $\omega_q$ needs to be rapidly modulated between two values $\omega_q/2\pi=\omega_c+\delta h\cdot h_0$ for $U_+$, while $\omega_q/2\pi=\omega_c-\delta h\cdot h_0$ for $U_-$, where $\omega_c$ is the common qubit frequency.  For the frequency-tunable transmon qubit, the relationship between $\omega_q$ and the amplitude of Z pulse $z$ is
    \begin{equation}
        \omega_q=\sqrt{8E_{JJ}E_C|\cos(k z+b)|}-E_C,
    \end{equation}
    where $E_{JJ}$ denotes the Josephson energy, $E_C$  is the charging energy, and $kz+b=\Phi_{\textrm{ext}}/\Phi_0$ with the weak external flux $\Phi_{\textrm{ext}}$. The RMD pulse consists of a series of square Z waves with duration of $T$. Owing to the constrained sampling rate of the DAC, both the falling and rising edge durations are inherently limited to a minimum of about $0.5$~ns. In our experiments, these square waves are substituted to trapezoidal waves with edges of $0.5$~ns. Additionally, we present the waveform captured by the oscilloscope in Supplementary Fig.~\ref{fig_rmd_pulse_exp}.

	\begin{figure}[h]
		\centering
		\includegraphics[width=0.7\linewidth]{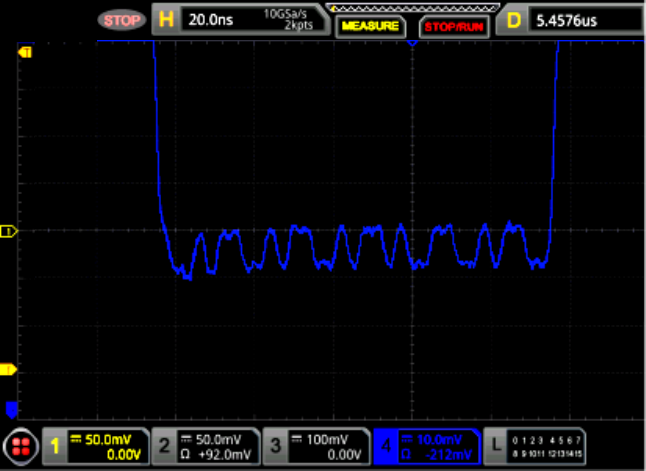}
		\caption{\textbf{Waveforms obtained from oscilloscope measurements.}} 
		\label{fig_rmd_pulse_exp}
	\end{figure}
	
	\begin{figure}[t]
		\centering
		\includegraphics[width=0.8\linewidth]{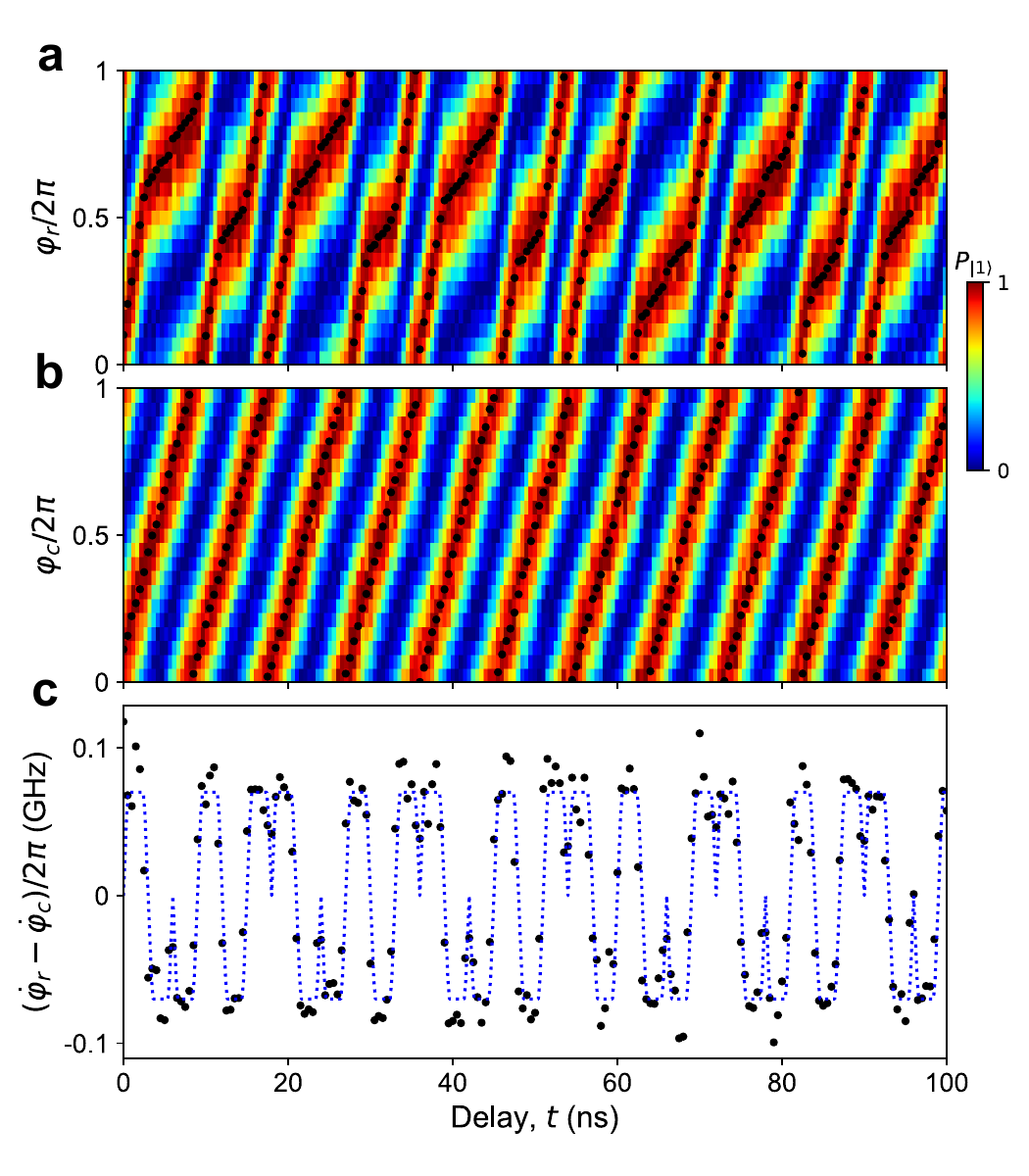}
		\caption{\textbf{Characterization of the random multipolar driving (RMD) pulse.} \textbf{a}, The measured phase $\varphi_r$, when employing RMD pulse, as a fucntion of RMD pulse duration. \textbf{b}, The measured phase, $\varphi_c$, obtained by applying rectangle Z pulse, versus the pulse duration. \textbf{c}, The qubit frequency, derived by differentiating $\varphi_r-\varphi_c$ with respect to time, as a function of the delay. The dashed curve represents the engineered RMD sequence, with the characteristic timescale $T=3$~ns, $n=1$ and the amplitude of $\delta h\cdot h/2\pi=70$~MHz. Specially, the RMD sequence reads $\{U_+U_-U_-U_+U_-U_+\cdots\}$.}
		\label{RMDPulseshape}
	\end{figure}
	 
	 To further characterize the RMD pulse, we measure the dynamical phase induced during its operation. Initially, the qubit is prepared at its idle point and excited using an $X/2$ gate. Then, the qubit is biased to the working point using the RMD pulse and the flat pulse following a delay, respectively. After turning off all Z pulses to tune the qubit back to its idle point, we apply another rotation $R_{\phi}(\frac{\pi}{2})$, with $\phi$ ranging from $0$ to $2\pi$. The population of $\ket{1}$ state reaches its maximum only when $\phi$ compensates for the accumulated dynamical phase. In Supplementary Figs.~\ref{RMDPulseshape}\textbf{a} and \textbf{b}, we show the accumulated phase in the RMD case
	\begin{equation}
	    \varphi_r(t)=\int_0^t\!\mathrm{d}t\;(\omega_r (t)-\omega_{\textrm{idle}}),
	\end{equation}
	where $\omega_r(t)$ is the qubit frequency under RMD, and the phase in the flat case
	\begin{equation}
	    \varphi_c(t)=\int_0^t\!\mathrm{d}t\;(\omega_c-\omega_{\textrm{idle}}).
	\end{equation}
	The RMD sequence is precisely recovered by differentiating $\varphi_r-\varphi_c$ with respect to time
	\begin{equation}
	    \omega_r(t)-\omega_c=\frac{\mathrm{d}}{\mathrm{d}t}(\varphi_r(t)-\varphi_c(t)).
	\end{equation}
	The result, as depicted in Supplementary Fig.~\ref{RMDPulseshape}\textbf{c}, is well in accordance with the engineered sequence.
	
	\subsection{Timing by Floquet engineering}
	
	Another challenge in achieving precise RMD control lies in the timing misalignment between adjacent qubits, when applying rapid pulse with the short period less than $5$~ns. 
	
	Here, we employ Floquet engineering~\cite{Shi2023} to synchronize the timing sequence of the flux bias pulses between qubit-qubit pairs. As shown in Supplementary Fig.~\ref{FloquetTiming}\textbf{a}, the Floquet sequence $\{U_-U_+U_-U_+U_-U_+-\cdots\}$, with the same characteristic time $T$, is simultaneously applied to two adjacent qubits. The coupling strength is dynamically modulated according to the degree of timing misalignment between the qubits. To mimic the the modulation dynamics, we use a sinusoidal waveform ($j=1, 2$)
	\begin{equation}
	    \omega_{j}(t)=\omega_c+A\sin(\mu t+\varphi_j),
	\end{equation}
	where $A=\delta h\cdot h$ is the driving amplitude, $\mu=\pi/T$ is the driving frequency, and $\varphi_j$ characterizes the timing offset. The Hamiltonian reads
	\begin{equation}
	    \hat{H}=-\frac{1}{2}\omega_1(t)\sigma_1^z-\frac{1}{2}\omega_2(t)\sigma_2^z+g(\sigma_1^+\sigma_2^-+\mathrm{H.c.}),
	\end{equation}
	with Pauli matrices $\sigma_{x,y,z}$, $\sigma_{\pm}=(\sigma_x\pm\mathrm{i}\sigma_y)/2$, and the direct coupling strength $g$. By applying the unitary transformation
	\begin{equation}
	    \hat{U}=\textrm{exp}\left\{\mathrm{i}\int_0^t\!\mathrm{d}t\;\left[\omega_1(t)+\omega_2(t)\right]\right\}=\textrm{exp}\left\{-\frac{1}{2}\mathrm{i}\sum_{j}\left[\omega_ct+\frac{A}{\mu}\cos\left(\mu t+\varphi_j\right)\right]\sigma_j^z\right\},
	\end{equation}
	the effective Hamiltonian can be calculated as
	\begin{subequations}
	\begin{align}
	    \hat{H}'=&\hat{U}\hat{H}\hat{U}^{\dagger}+\mathrm{i}\partial_t\hat{U}\hat{U}^{\dagger}\\
	    =&g\exp\left\{\mathrm{i}\frac{A}{\mu}\left[\cos(\mu t+\varphi_1)-\cos(\mu t+\varphi_2)\right]\right\}\sigma_1^+\sigma_2^-+\mathrm{H.c.}\\
	    =&g\exp\left\{2\mathrm{i}\frac{A}{\mu}\sin\left(\frac{\Delta\varphi}{2}\right)\sin\left(\mu t+\frac{\varphi_1+\varphi_2}{2}\right)\right\}\sigma_1^+\sigma_2^-+\mathrm{H.c.}\\
	    =&g\sum_{n=-\infty}^{\infty}J_n\left(2\frac{A}{\mu}\sin\left(\frac{\Delta\varphi}{2}\right)\right)\mathrm{e}^{\mathrm{i}n\theta}\sigma_1^+\sigma_2^-+\mathrm{H.c.},
	\end{align}
	\end{subequations}
	where $\Delta\varphi=\varphi_2-\varphi_1$ is the phase difference, $\theta=\mu t+(\varphi_1+\varphi_2)/2$, $J_n(\cdot)$ is the $n$-order Bessel function, and the last line is from Jacobi-Anger series $\mathrm{e}^{\mathrm{i}x\sin\theta}=\sum_nJ_n(x)\mathrm{e}^{\mathrm{i}n\theta}$. Under rotating-wave approximation (RWA), when the condition $\mu\gg g$ is satisfied, only the lowest frequency term $n=0$ remains, leading to the effective coupling strength
	\begin{equation}
	    g^{\textrm{eff}}=g J_0(2\frac{A}{\mu}\sin(\frac{\Delta\varphi}{2})).
	\end{equation}
	
    The effective coupling strength reaches its maximum value $g$, if and only if the phase difference satisfies 
    $\Delta\varphi=2m\pi$ with an arbitrary integer $m$. This relationship enables us to achieve precise timing alignment by systematically characterizing the coupling strength as a function of the phase difference $\Delta\varphi$ or, equivalently, the timing offset $\Delta t$ in the Floquet sequence implementation. A typical result is shown in Supplementary Fig.~\ref{FloquetTiming}\textbf{b} and \textbf{c}.
    
    \begin{figure}[t]
		\centering
		\includegraphics[width=0.85\linewidth]{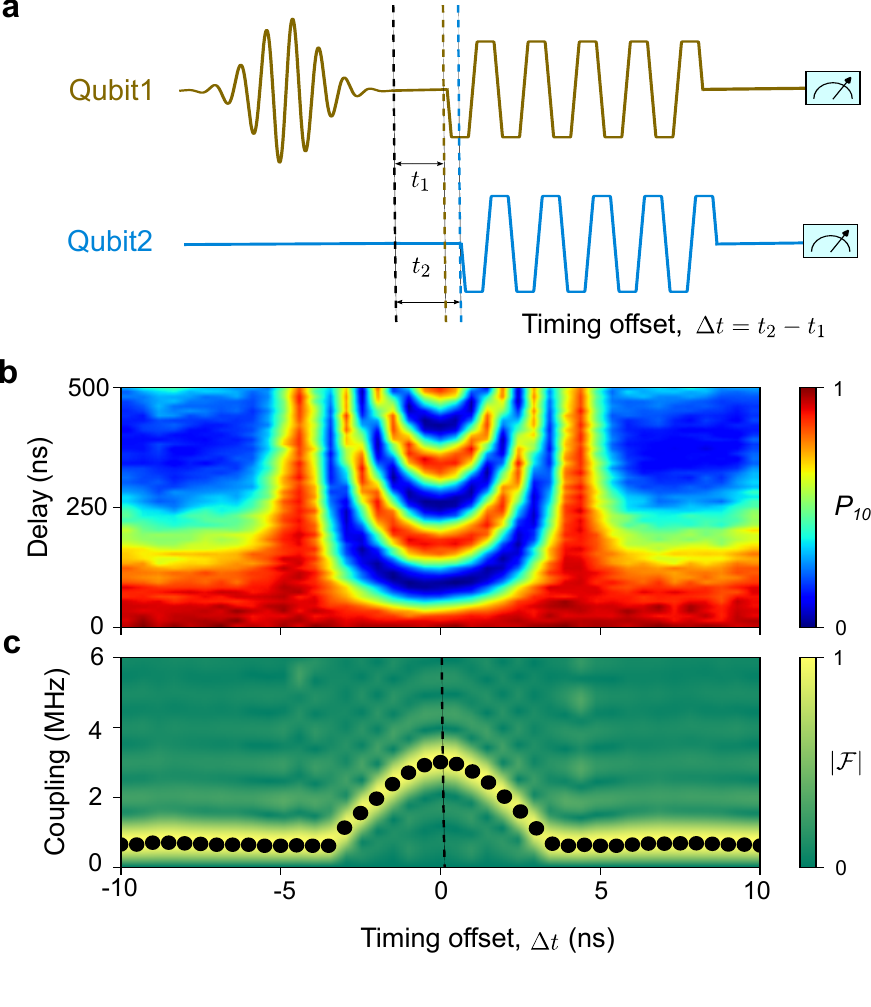}
		\caption{\textbf{ Synchronization of the flux bias pulses between qubit-qubit pairs by Floquet engineering.} \textbf{a}, Experimental pulse sequences for calibrating the timing between the flux bias pulses of a qubit-qubit pair. The qubit is excited to the state $\ket{1}$ using a $\pi$ pulse, followed by the measurement of the coupling strength as a function of the timing offset.  \textbf{b}, The experimentally measured $\ket{10}$ probability, of the qubit versus the timing offset $\Delta t$. \textbf{c}, Fourier spectrum of the data presented in \textbf{b}. The characteristic time $T$ is chosen as $5$~ns.}
		\label{FloquetTiming}
	\end{figure}
    \newpage
    
    \section{Supplementary Data}
    \subsection{Dynamics of imbalance in systems with different sizes}
	\begin{figure}[H]
		\centering
		\includegraphics[width=0.78\linewidth]{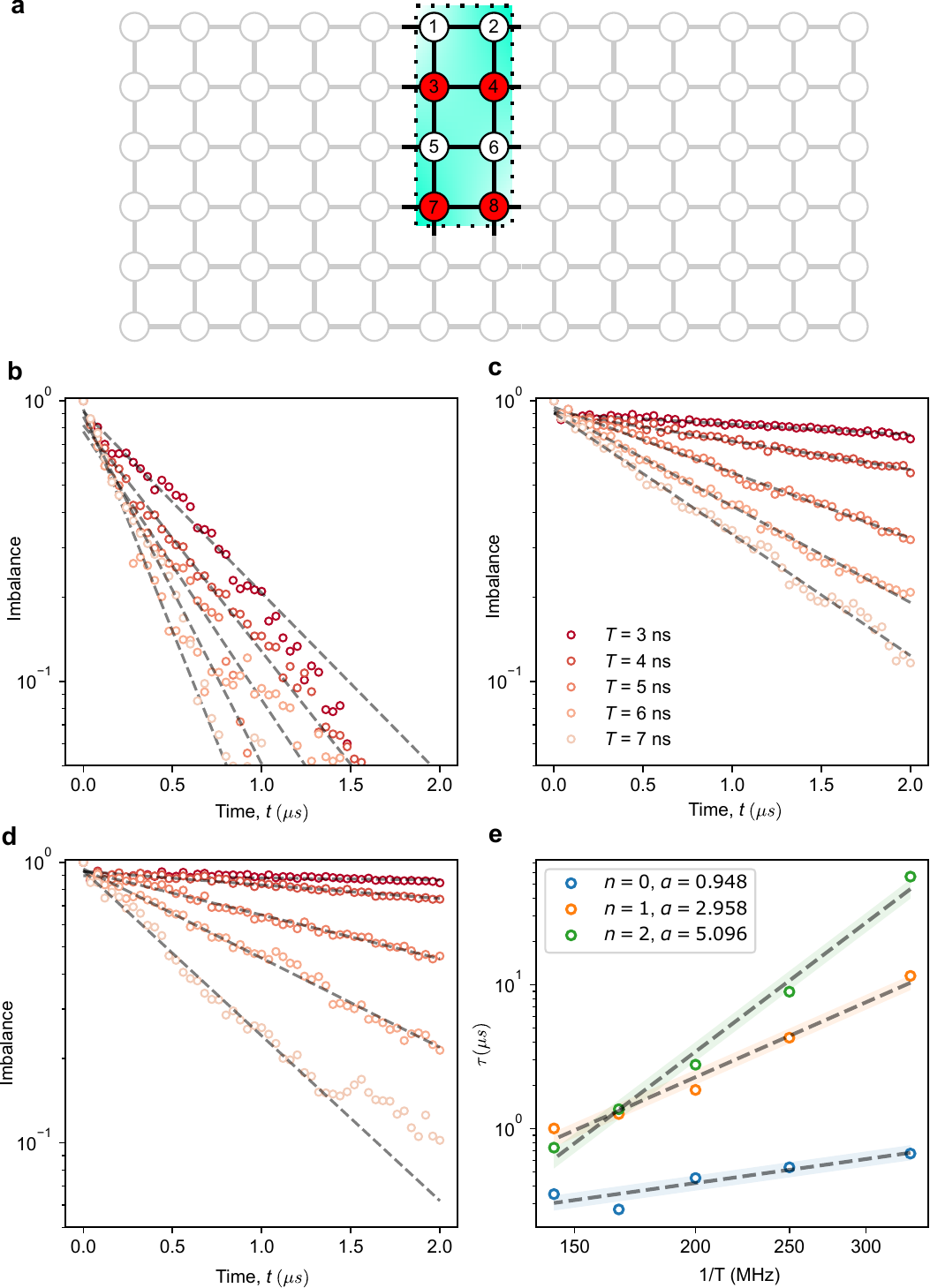}
		\caption{\textbf{The dynamics of imbalance in the system of size $2\times 4$.} \textbf{a}, Spatial configuration of the $2\times 4$ system  within the processor. \textbf{b-d}, Time evolution of the imbalance for driving periods $T$ ranging from 3~ns to 7~ns, corresponding to $n=0$, $n=1$, $n=2$, individually. Markers represent experimental data, while dashed lines denote fitted curves used to extract the decay rate $-1/\tau_I$, where $\tau_I$ is the characteristic decay time. \textbf{e}, Power-law scaling of $\tau_I$, presented on a log-log scale. The scaling exponents, $\alpha(n=0)=0.948, \alpha(n=1)=2.958, \alpha(n=2)=5.096$, follow the theoretical prediction, $\alpha=2n+1$.
		}
		
		\label{imbalance8q_1}
	\end{figure}

	\begin{figure}[H]
		\centering
		\includegraphics[width=0.8\linewidth]{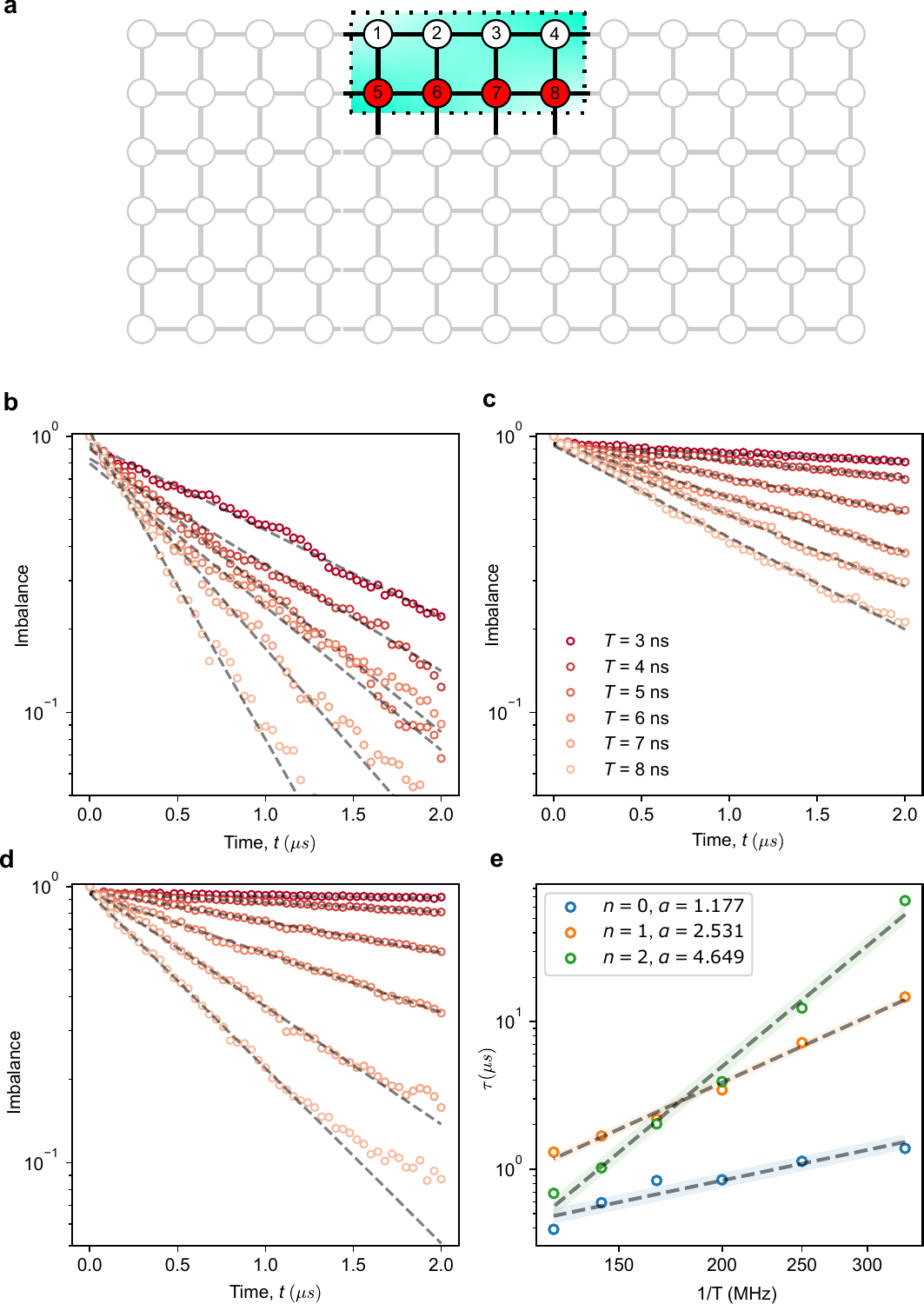}
		\caption{\textbf{The dynamics of imbalance in the system of size $4\times 2$.} \textbf{a}, Spatial configuration of the $4\times 2$ system  within the processor. \textbf{b-d}, Time evolution of the imbalance for driving periods $T$ ranging from 3~ns to 8~ns, corresponding to $n=0$, $n=1$, $n=2$, individually. Markers represent experimental data, while dashed lines denote fitted curves used to extract the decay rate $-1/\tau_I$, where $\tau_I$ is the characteristic decay time.
		\textbf{e}, Power-law scaling of $\tau_I$ in a log-log scale, with extracted exponents  $\alpha(n=0)=1.177, \alpha(n=1)=2.531, \alpha(n=2)=4.649$. 
		}
		\label{imbalance8q_2}
	\end{figure}
    
%     \begin{figure}[h]
% 		\centering
% 		\includegraphics[width=0.9\linewidth]{imbalance24q.pdf}
% 		\caption{\textbf{Imbalance of size $6\times 4$}.
% 		 (a) N=0, imbalance of different $U_{+}(U_{-})$ periods $T$  from \SI{3}{\nano\second} to \SI{8}{\nano\second}. 
% 		(b) N=1, imbalance of different $U_{+}(U_{-})$ periods $T$ from \SI{3}{\nano\second} to \SI{8}{\nano\second}.
% 		(c) N=2, imbalance of different $U_{+}(U_{-})$ periods $T$ from \SI{3}{\nano\second} to \SI{8}{\nano\second}.
% 		(d) Use linear fitting to get $2n+1$ power law of $\tau$ v.s. $1/T$. The slope $\alpha_{n=0}=0.925, \alpha_{n=1}=2.902, \alpha_{n=2}=4.002$.
% 		} 
% 		\label{imbalance24q}
% 	\end{figure}
	
	\begin{figure}[H]
		\centering
		\includegraphics[width=0.85\linewidth]{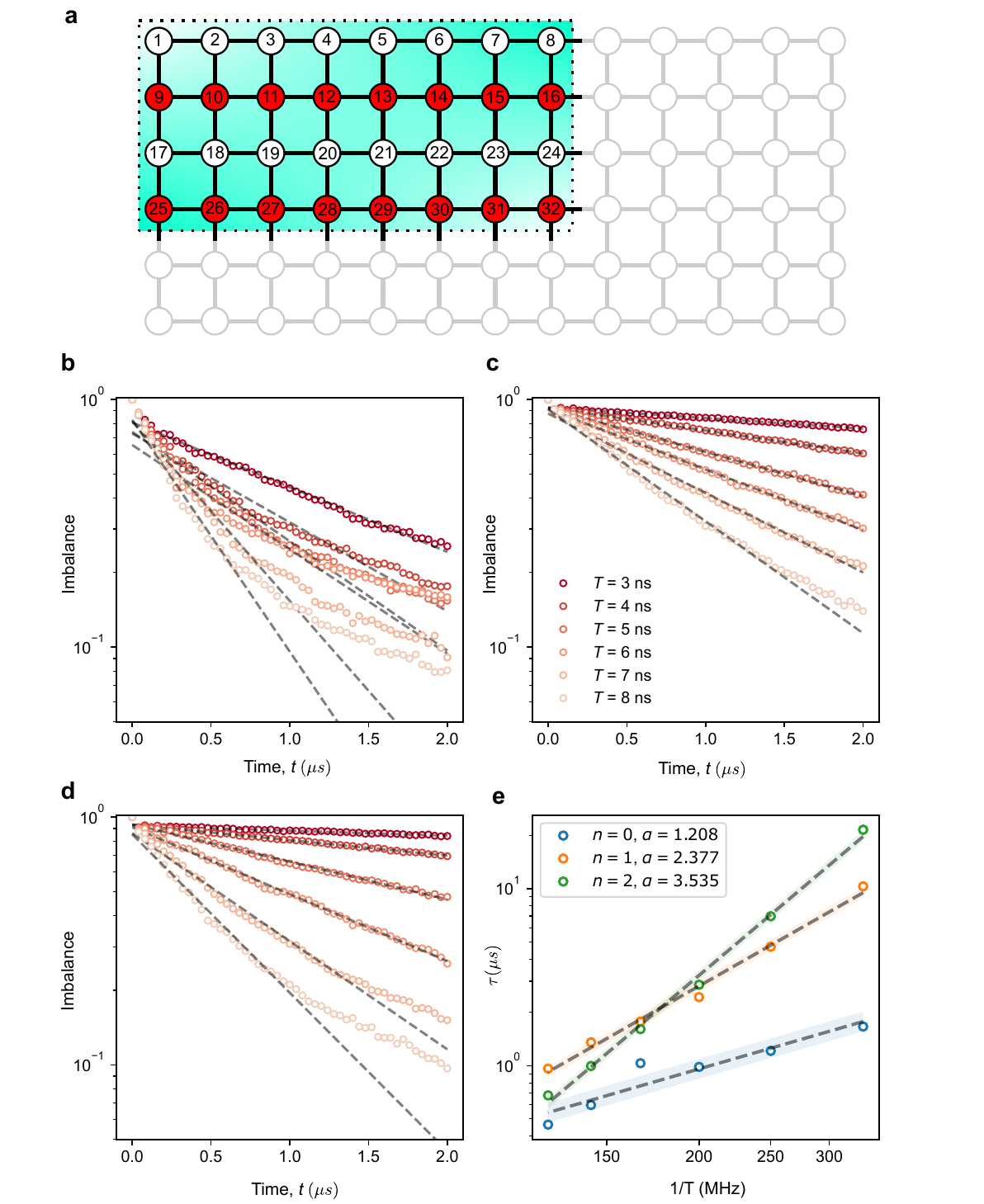}
		\caption{\textbf{The dynamics of imbalance in the system of size $8\times 4$.} \textbf{a}, Spatial configuration of the $8\times 4$ system  within the processor. \textbf{b-d}, Time evolution of the imbalance for driving periods $T$ ranging from 3~ns to 8~ns, corresponding to $n=0$, $n=1$, $n=2$, individually. 
		\textbf{e}, Power-law scaling of the imbalance decay time, presented in a log-log scale, with extracted exponents  $\alpha(n=0)=1.208, \alpha(n=1)=2.377, \alpha(n=2)=3.535$.
		}
		\label{imbalance32q}
	\end{figure}
	
	\begin{figure}[H]
		\centering
		\includegraphics[width=0.8\linewidth]{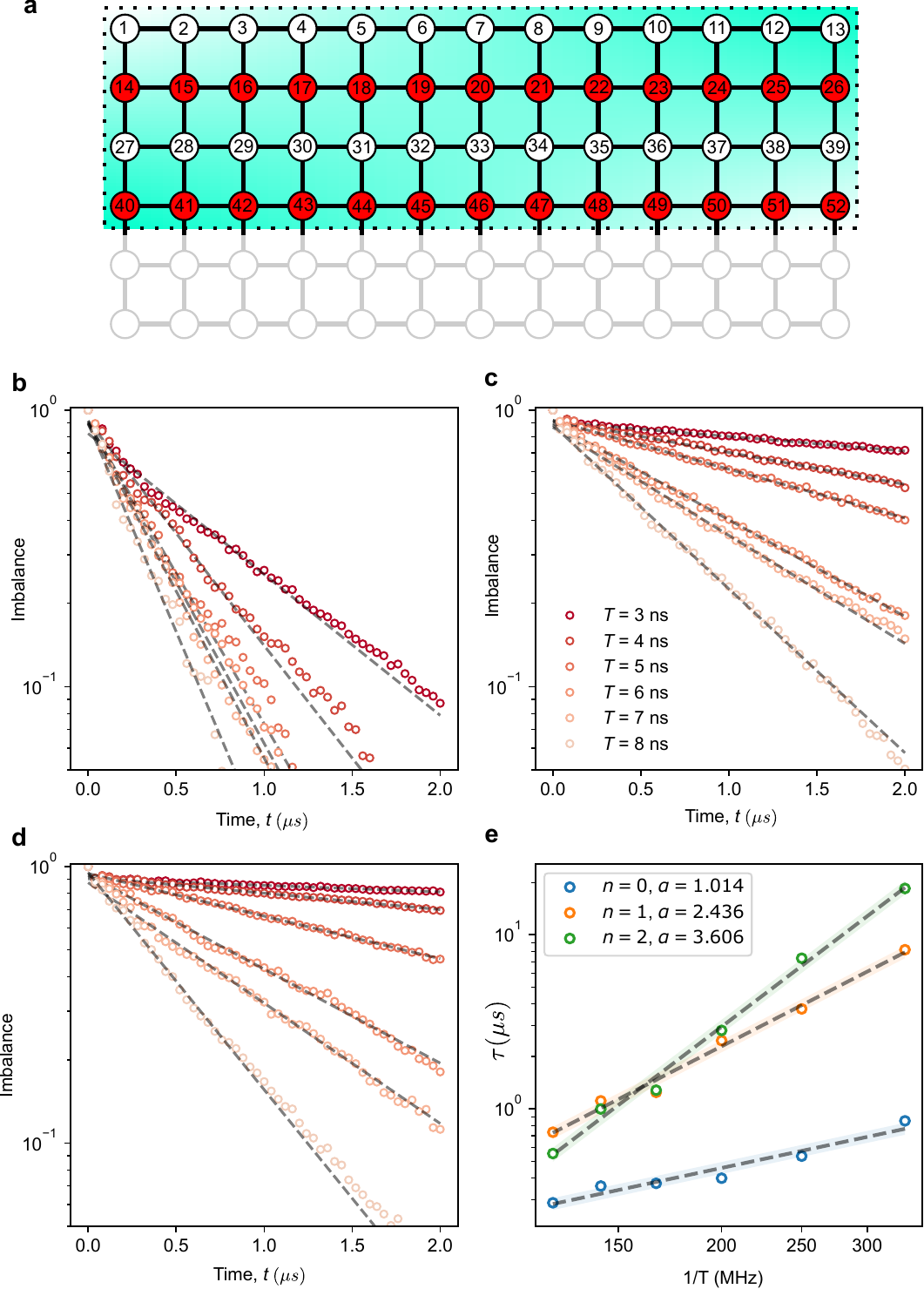}
		\caption{\textbf{The dynamics of imbalance in the system of size $13\times 4$.} \textbf{a}, Spatial configuration of the $13\times 4$ system  within the processor. \textbf{b-d}, Time evolution of the imbalance for driving periods $T$ ranging from 3~ns to 8~ns, corresponding to $n=0$, $n=1$, $n=2$, individually. Markers represent experimental data, while dashed lines denote fitted curves used to extract the decay rate $-1/\tau_I$, where $\tau_I$ is the characteristic decay time.
		\textbf{e}, Power-law scaling of $\tau_I$, presented on a log-log scale, with extracted exponents  $\alpha(n=0)=1.014, \alpha(n=1)=2.436, \alpha(n=2)=3.606$. 
		}
		\label{imbalance52q}
	\end{figure}
    
    \begin{figure}[H]
		\centering
		\includegraphics[width=0.8\linewidth]{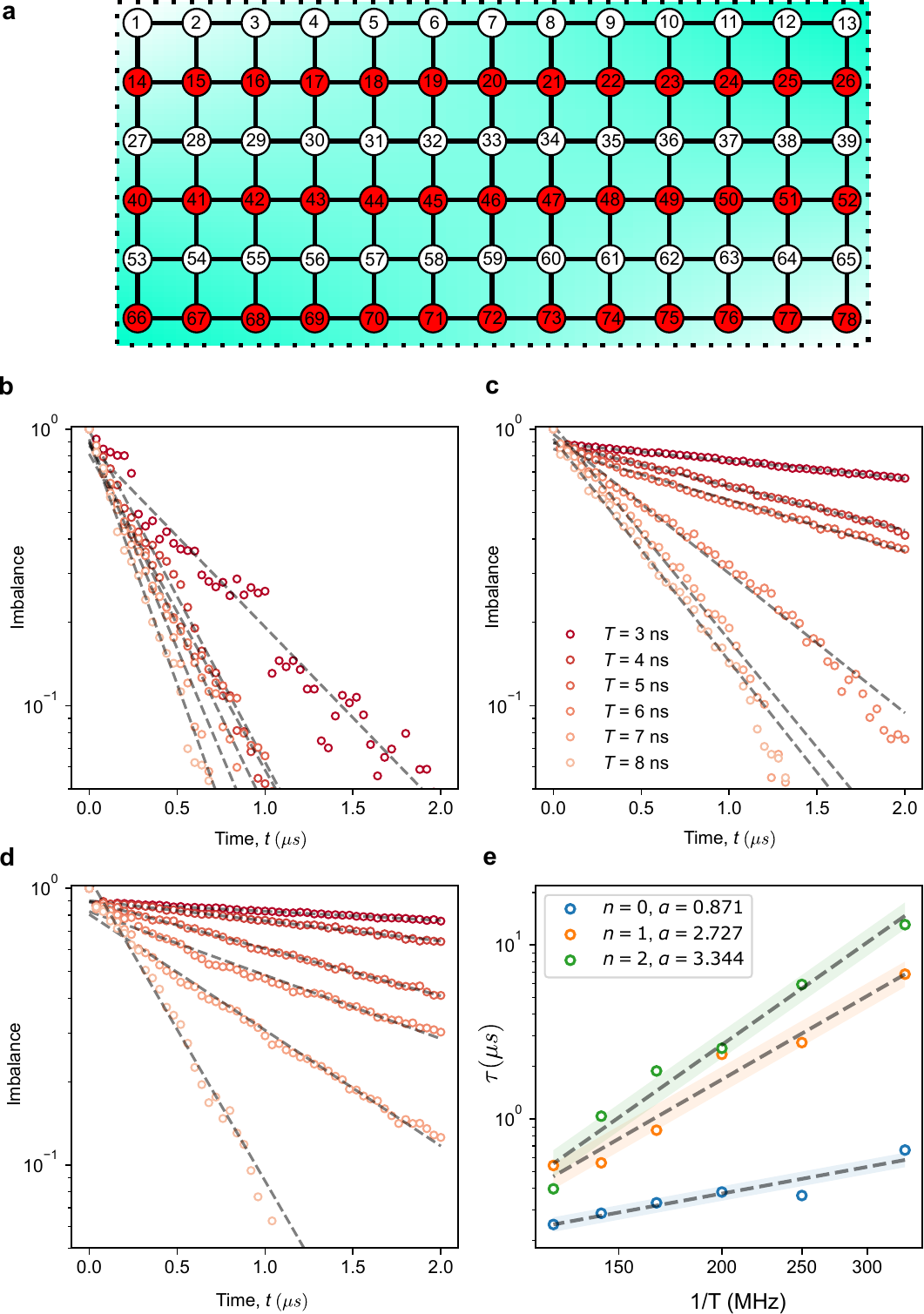}
		\caption{\textbf{The dynamics of imbalance in the system of size $13\times 6$.} \textbf{a}, Spatial configuration of the $13\times 6$ system  within the processor. \textbf{b-d}, Time evolution of the imbalance for driving periods $T$ ranging from 3~ns to 8~ns, corresponding to $n=0$, $n=1$, $n=2$, individually. Markers represent experimental data, while dashed lines denote fitted curves used to extract the decay rate $-1/\tau_I$, where $\tau_I$ is the characteristic decay time.
		\textbf{e}, Power-law scaling of $\tau_I$, presented on a log-log scale, with extracted exponents  $\alpha(n=0)=0.871, \alpha(n=1)=2.727, \alpha(n=2)=3.344$.
		} 
		\label{imbalance78q}
	\end{figure}
	
    \subsection{Dynamics of entropy in systems with different sizes}
    
    \begin{figure}[h]
		\centering
		\includegraphics[width=0.85\linewidth]{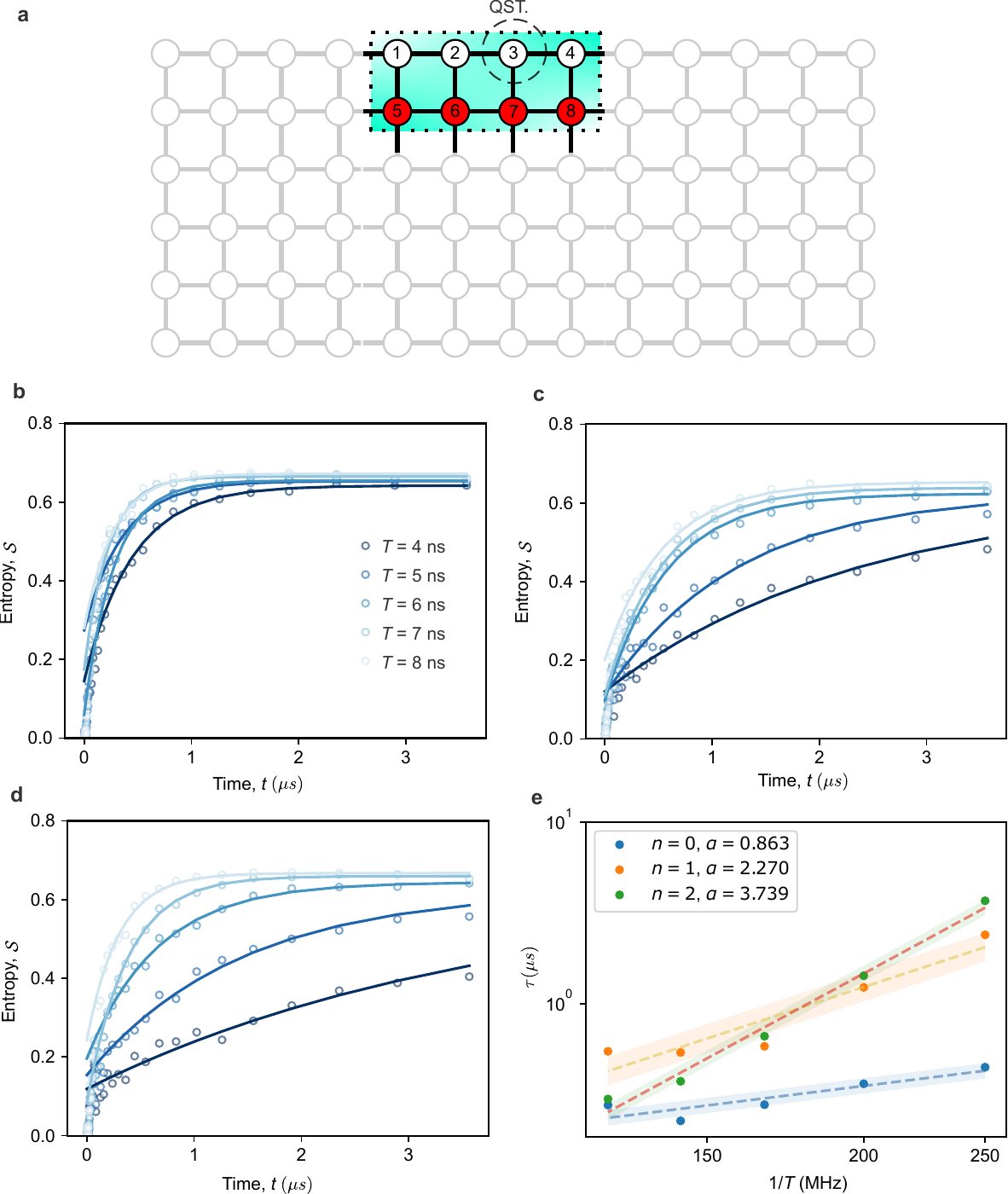}
		\caption{\textbf{The dynamics of entropy in the system of size $4\times 2$.} \textbf{a}, Spatial configuration of the $4\times 2$ system and the subsystem for performing quantum state tomography within the processor. \textbf{b-d}, Time evolution of the subsystem entanglement entropy for driving periods $T$ ranging from 4~ns to 8~ns, corresponding to $n=0$, $n=1$, $n=2$, individually. 
		\textbf{e}, Power-law scaling of the prethermal lifetime presented on a log-log scale, with extracted exponents $\alpha(n=0)=0.863, \alpha(n=1)=2.270, \alpha(n=2)=3.739$. The curves in \textbf{b}-\textbf{d} are the fitted results using the form $S\sim S_M(1-e^{-t/\tau_S})$.
		} 
		\label{entropy_8Q}
	\end{figure}
    
    \begin{figure}[h]
		\centering
		\includegraphics[width=0.85\linewidth]{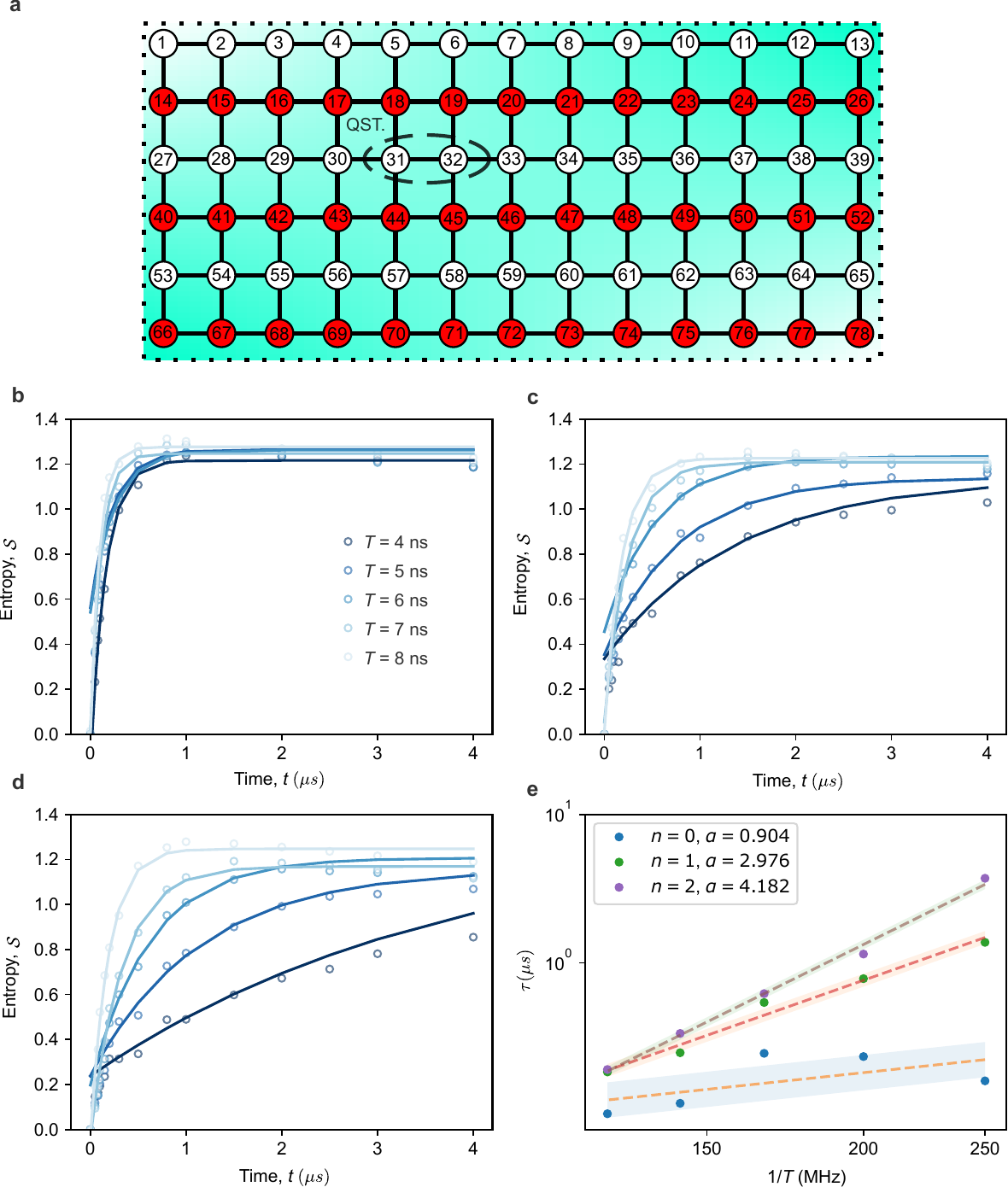}
		\caption{\textbf{The dynamics of entropy in the system of size $13\times 6$.} \textbf{a}, Spatial configuration of the $13\times 6$ system and the subsystem for performing quantum state tomography (QST) within the processor. \textbf{b-d}, Time evolution of the subsystem entanglement entropy for driving periods $T$ ranging from 4~ns to 8~ns, corresponding to $n=0$, $n=1$, $n=2$, individually. 
		\textbf{e}, Power-law scaling of the prethermal lifetime in a log-log scale, with extracted exponents $\alpha(n=0)=0.904, \alpha(n=1)=2.976, \alpha(n=2)=4.182$. The curves in \textbf{b}-\textbf{d} are the fitted results using the form $S\sim S_M(1-e^{-t/\tau_S})$.
		}
		\label{entropy_78Q}
	\end{figure}

%
% ****** End of file apssamp.tex ******

\end{document}